\documentclass{aa}  

\usepackage{graphicx}
\usepackage{txfonts}

\usepackage{ulem}
\usepackage{xcolor}
\usepackage{stackengine}
\usepackage{bm}
\usepackage{orcidlink}
\usepackage[amsmath,thref,hyperref]{ntheorem}
\usepackage{placeins}
\usepackage{float}
\usepackage{hyperref}
\hypersetup{
    colorlinks=true,
    linkcolor=blue,
    urlcolor=blue,
    citecolor=blue,
    pdfauthor=author
    }

\newcommand{\degree}{^{\circ}}
\newcommand{\absrm }{{\rm |RM|}}
\newcommand{\Ne}    {n_{\rm e}}
\newcommand{\Si}    {S_{I}}
\newcommand{\Sp}    {S_{\rm p}}
\newcommand{\Ii}    {I_{I}}
\newcommand{\Ip}    {I_{\rm p}}
\newcommand{\Mp}    {m_{\rm p}}
\newcommand{\chiobs}{\chi_{\rm obs}}
\newcommand{\Tb}    {T_{\rm b}}
\newcommand{\Teq}   {T_{\rm eq}}
\newcommand{\vang}  {\theta_{\rm v}}
\newcommand\angdot[1][\circ]{%
  \stackengine{0pt}{.}{${}^{\mathrm{#1}}$}{O}{l}{F}{F}{L}}
\begin{document}

    \title{Polarization flare of 3C~454.3 in millimeter wavelengths seen from decadal polarimetric observations}

    \author{
    Hyeon-Woo Jeong\inst{1,2}$^{\orcidlink{0009-0005-7629-8450}}$
    \and
    Sang-Sung Lee\inst{1,2}\fnmsep\thanks{Corresponding author. e-mail$\colon$sslee@kasi.re.kr}$^{\orcidlink{0000-0002-6269-594X}}$
    \and
    Sincheol Kang\inst{2}$^{\orcidlink{0000-0002-0112-4836}}$
    \and
    Minchul Kam\inst{3,4}$^{\orcidlink{0000-0001-9799-765X}}$
    \and
    Sanghyun Kim\inst{1,2}$^{\orcidlink{0000-0001-7556-8504}}$
    \and
    Whee Yeon Cheong\inst{1,2}$^{\orcidlink{0009-0002-1871-5824}}$
    \and
    Do-Young Byun\inst{1,2}$^{\orcidlink{0000-0003-1157-4109}}$
    \and
    Chanwoo Song\inst{1,2}$^{\orcidlink{0009-0003-8767-7080}}$
    \and
    Sascha Trippe\inst{4,5}$^{\orcidlink{0000-0003-0465-1559}}$
    }

    \institute{
    Astronomy and Space Science, University of Science and Technology, 217 Gajeong-ro, Yuseong-gu, Daejeon 34113, Republic of Korea
    \and
    Korea Astronomy and Space Science Institute, 776 Daedeok-daero, Yuseong-gu, Daejeon 34055, Republic of Korea
    \and
    Institute of Astronomy and Astrophysics, Academia Sinica, P.O. Box 23-141, Taipei 10617, Taiwan
    \and
    Department of Physics and Astronomy, Seoul National University, Gwanak-gu, Seoul 08826, Republic of Korea
    \and
    SNU Astronomy Research Center, Seoul National University, Gwanak-gu, Seoul 08826, Korea
    }

    \date{Received ???; accepted ???}

 
   \abstract
   {The blazar 3C~454.3~($z=0.859$) has been extensively investigated using multi-wavelength high-resolution polarization studies, showing polarization variations on the milli-arcsecond (mas) scale.}
   {This study investigates polarimetric characteristics of the blazar 3C~454.3 at 22--129~GHz using decadal~(2011--2022) data sets. In addition, we also delve into the origin of the polarization flare observed in 2019.}
   {The corresponding data sets were obtained from the single-dish mode observations of the Korean VLBI Network~(KVN) and the 43-GHz Very Long Baseline Array~(VLBA). Using those data, we compared the consistency of the measurements between milli-arcsecond and arcsecond scales. The Faraday rotation measure~(RM) values were obtained via two approaches, model fitting to a linear function in all frequency ranges, and calculation from adjacent frequency pairs.}
   {We found that the linear polarization angle is preferred to be $\sim100\degree$ when the source is highly polarized, for example, during a flare. At 43~GHz, we found that the polarized emission at scales of mas and arcsecond is consistent when we compare its flux density and polarization angle. The ratio of quasi-simultaneously measured~(within a week) polarized flux density is $1.02\pm0.07$~(i.e., $\Delta S_{\rm p}/S_{\rm p}\approx2\%$), and the polarization angles display similar rotation. These suggest that the extended jet beyond the scale of VLBA 43~GHz has a negligible convolution effect on the polarization angle from the KVN. We found an interesting, notable flaring event in the KVN single-dish data from the polarized emission in 2019 in the frequency range of 22--129~GHz. During the flare, the observed polarization angles~($\chiobs$) rotate from $\sim150\degree$ to $\sim100\degree$ at all frequencies with a chromatic polarization degree~($\Mp$).}
   {Based on the observed $\Mp$ and $\chiobs$, and also the Faraday rotation measure, we suggest that the polarization flare in 2019 is attributed to the shock--shock interaction in the stationary jet region. The change in the viewing angle of the jet alone is insufficient to describe the increase in brightness temperature, indicating the presence of source intrinsic processes, such as, e.g., particle acceleration.}


    \titlerunning{Polarization flare of 3C~454.3 in millimeter wavelengths}
    \maketitle
%

\section{Introduction}
\label{sec:intro}
    The blazar is a subclass of Active Galactic Nucleus~(AGN), in which a relativistic jet is inclined with a small viewing angle between the jet and the line-of-sight. The small viewing angle of the jet enhances the Doppler boosting effects, leading to rapid variability. The jet emits radiation over the entire range of electromagnetic waves, which is dominated by synchrotron radiation from radio to optical bands, and radiation by inverse-Compton scattering in higher energy bands~(e.g., $\gamma$-ray). Synchrotron radiation is emitted by relativistic electrons gyrating around magnetic field lines in the jet.
    
    The synchrotron radiation from the relativistic jets is most likely polarized, with its polarized properties strongly constrained by magnetic field environments in the jet. If the field lines in the jets are well-ordered and strong, one may observe a high degree of linear polarization~($\Mp$) and high polarized flux density~($\Sp$). Theoretically, for an ideal case~(i.e., a perfectly ordered magnetic field), $\Mp$ reaching up to $\sim75\%$ is feasible for electrons having power-law energy distribution in an optically thin~(e.g., $\alpha=-1$, where $S \propto \nu^{\alpha}$) region~\citep{rybicki1979, review_trippe2014}. Meanwhile, for an optically thick region, the observed $\Mp$ is expected to decrease, e.g., only a few percent of $\Mp$.
    
    Also, observational results have identified polarized emission from AGN, and both $\Mp$ and the polarization angle ($\chi$) seem to give information about the magnetic field lines in a jet. Several studies have reported the results of polarization observations in both radio and optical bands for AGNs and found highly polarized emission reaching a few tens of percent in polarization degree~\citep{jorstad2005, jorstad2007, pushkarev2023}. As the optically thick region is believed to have low $\Mp$ even with perfectly ordered magnetic field lines, this implies an optically thin polarized emission from some AGNs at high radio frequency~$(\rm e.g.,~\gtrsim86~GHz)$, since optical wavelengths are optically thin by default. \citet{angelakis2016} suggested a model discriminating low- and high-synchrotron peak~(LSP and HSP, respectively) blazars based on the optical polarimetry~\citep[see also][]{blinov2019}. In their work, in optical wavelengths, it seems that LSP blazars exhibit highly polarized emission with a random $\chi$ distribution and opposite characteristics~(i.e., a low degree of polarization with a specific $\chi$) for HSP blazars. 
    
    Recently, at shorter wavelengths, the Imaging X-ray Polarimetric Explorer~(IXPE) has been observing polarized emission X-rays from celestial objects, including AGNs, and has identified jet models of the AGNs by comparison with radio and optical polarization behavior~\citep{gesu2022, marinucci2022, liodakis2022, middei2023_bllac, peirson2023, middei2023_1553, kim2024}. Among the X-ray studies, the highest measurement in $\Mp$ is about $22\%$. However, we note that the polarization degrees listed above are not representative values at each observing frequency since the samples and observing periods are different from each other.
    
    Although the observed polarization angle $\chiobs$ most likely provides us with information about the orientation of magnetic field lines in the jet, the Faraday effect contaminates the intrinsic polarized emission when the emission propagates through an ionized medium with magnetic field lines~\citep{burn1966, sokoloff1998}. This leads to a rotation of $\chi$, which is proportional to the squared wavelength, $\Delta \chi \propto {\rm RM}\,\lambda^{2}$, where $\Delta \chi = \chiobs - \chi_{0}$, and $\chi_{0}$ is the intrinsic polarization angle. The observed polarization angle can then be written in the form of ${\chiobs} = \chi_{0} + {\rm RM}\,\lambda^{2}$, where $\chiobs$ is the observed polarization angle, and $\lambda$ is the observing wavelength. RM is the Faraday rotation measure, indicating the amount of rotation in the polarization angle by a medium in units of $\rm rad\,m^{-2}$. Therefore, the Faraday rotation is larger at longer wavelengths. Based on this expression, one can calculate the RM using the following: ${\rm RM} = (\chi_{1}-\chi_{2})/({\lambda_{1}}^{2}-{\lambda_{2}}^{2})$, where $\chi_{1}$ and $\chi_{2}$ are the observed polarization angles at wavelengths $\lambda_{1}$ and $\lambda_{2}$, respectively. For a cosmological source at redshift $z$, RM is defined as$\colon$
    \begin{equation}
    \begin{aligned}
    {\rm RM} = 
    \frac {e^{3}} {2 \pi {m_{\rm e}}^{2} {c}^{4}}\, \frac{1} {(1+z)^{2}}
    \int \Ne(l)\,B_{\parallel}(l)\, dl,
    \label{eq:RM}
    \end{aligned}
    \end{equation}
    where $m_{\rm e}$ and $e$ are the mass and charge of an electron, $c$ is the speed of light, $\Ne$ is the electron number density of a Faraday screening medium, $B_{\parallel}$ is the magnetic field strength along the line-of-sight, and $l$ is the path length through the medium where the Faraday rotation takes place.
    
    In millimeter wavelengths, the Institute for Radio Astronomy in the Millimeter Range~(IRAM) 30~m radio telescope has monitored 37 AGNs at 86 and 229~GHz~\citep{agudo2018_polami1} in dual-polarization mode to investigate the linear polarization variability of the AGNs~\citep{agudo2018_polami3}. In their work, they found faster variability and larger fractional linear polarization at 229~GHz than at 86~GHz.
    
    The sign of RM depends on $B_{\parallel}$ to the line-of-sight, as indicated in Equation \ref{eq:RM}. A few multi-wavelength polarization studies have revealed high absolute RM~($\absrm$) values of $\sim10^{5}~{\rm rad\,m^{-2}}$ in the observer's frame~\citep[e.g.,][]{kuo2014, park2018, agudo2018_polami3, hovatta2019}. In the Atacama Large Millimeter Array~(ALMA) observations, a higher order of $\absrm$~($\sim10^{7}~{\rm rad\,m^{-2}}$ in observer frame) was found from a gravitationally-lensed AGN called PKS~1830-211 at 250--300 GHz pair~\citep{marti-vibdal2015}. This was attributed to high $\Ne$ and/or strong $B_{\parallel}$ of the screening medium~(see Equation 1 in their work).
    
    The blazar 3C~454.3~($z=0.859$) has been extensively investigated in multi-wavelength high-resolution polarization studies. In very long baseline interferometry~(VLBI) observations at 15~GHz, the source was resolved with a widely extended jet structure beyond $\sim$10 milli-arcsecond~(mas) from the central engine, ejecting toward the northwest side~\citep{lister2018, lister2021, pushkarev2023}. At 43~GHz, two prominent regions have been identified over a decade$\colon$ the core and a quasi-stationary component~\citep[0.45--0.70~mas away from the core, hereafter, Region~C,][]{jorstad2017, weaver2022}. Typically, Region~C exhibits a higher polarization degree than that of the core with a stable polarization angle aligned to $\sim90\degree$~\citep[an east-west direction,][]{kemball1996, jorstad2005, jorstad2013, traianou2024}. Based on its observed properties, including re-brightening, decrease in the angular size of the jet, and spatial stationariness~\citep{gomez1999, jeong2023}, Region~C was suggested as a recollimation shock. For a downstream region, \citet{zamaninasab2013} found a large-scale helical magnetic field structure in 3C~454.3 by comparing the predictions of the helical jet model and observed polarimetric properties across the jet, using the multi-frequency Very Long Baseline Array~(VLBA\footnote{\url{https://science.nrao.edu/facilities/vlba}}) observations.
    
    Several studies were conducted to evaluate magnetic field strength in the jet of 3C~454.3. Based on the core-shift effect, the magnetic field strengths at the 43-GHz core and 1~pc from the central engine were found to be in 0.03--0.11~${\rm G}$ and 0.2--1.13~${\rm G}$, respectively, within 1-$\sigma$ uncertainty~\citep{pushkarev2012, kutkin2014, mohan2015, chamani2023}. In addition, \citet{jeong2023} estimated magnetic field strengths of the optically thick~($\tau=1$) surface in the jet via the synchrotron self-absorption~(SSA) effect. They found two distinctive SSA components, LSS and HSS~\citep[LSS$\colon$lower turnover frequency SSA spectrum and HSS$\colon$higher turnover frequency SSA spectrum; for more details, see Figure 4 and the text in][]{jeong2023}, which have turnover frequencies at low~($\lesssim30$~GHz) and high frequency~($\gtrsim60$~GHz) range. The estimated magnetic field strengths of the HSS were in the range of 0.1--7~${\rm mG}$. The LSS identified before the $\gamma$-ray flare in June 2014~\citep{gam_buson2014} corresponds to Region~C, and they found a magnetic dominance based on a higher SSA magnetic field strength than that in the equipartition condition.
    
    In this study, we investigate the decadal polarimetric characteristics of the source using the simultaneous multi-frequency~(22 to 129~GHz) single-dish observations from the Korean VLBI Network~(KVN\footnote{\url{https://radio.kasi.re.kr/kvn/main.php}}). We also employed the data from the 43-GHz VLBA, managed by the Boston University~(BU\footnote{\url{https://www.bu.edu/blazars/BEAM-ME.html}}) group. The obtained VLBA data were used to compare with the KVN data at 43~GHz and to resolve the source structure. The data sets used in this study have a time range from 2011--2022, and several $\gamma$-ray activities were reported in this period~\citep{gam_buson2014, gam_jorstad_2015, gam_ojha_2016, gam_panebianco_2022}. The observations included in this study will be described in Section~\ref{sec:observations}. The variability of the arcsecond-scale structure observed by the KVN single-dish and the variability of the mas-scale structures resolved by the VLBA are shown in Section~\ref{sec:lcs}. We investigated the Faraday rotation measure using the KVN observations in Section~\ref{sec:pa_rm}. In Section~\ref{sec:discussion}, we will discuss the observed features in the core and Region~C. The conclusions with summaries of this work will be in Section~\ref{sec:conclusions}.

\section{Observations and data reduction}
\label{sec:observations}
    \subsection{The KVN single-dish}
    \label{sec:obs_kvn}
        MOGABA~\citep[MOnitoring of GAmma-ray Bright Active galactic nuclei;][]{kang2015} is one of the polarimetric monitoring programs using the KVN on AGNs through the single-dish mode. Another monitoring program, PAGaN~\citep[Plasma-physics of Active Galactic Nuclei;][]{park2018}, also has been conducted for VLBI polarimetry, and their joint single-dish observation is used to calibrate the absolute polarization angle~\citep{kam2023}. 3C~454.3 is one of the target sources of MOGABA and PAGaN.
        
        This work used polarimetric monitoring data in the time periods 2011--2022~(MOGABA) and 2016--2022~(PAGaN) from the KVN 21-m radio telescopes at 22, 43, 86, and 129~GHz on 3C~454.3 in single-dish mode; the corresponding beam sizes are approximately 133, 68, 34 and 23 arcseconds. The KVN consists of three identical 21-m antennas located in Seoul~(KVN-Yonsei, KYS), Ulsan~(KVN-Ulsan, KUS), and Jeju~(KVN-Tamna, KTN) and has a unique capability of simultaneous four-frequency~(22, 43, 86, and 129~GHz) observation in VLBI mode. In a single-dish mode observation, two antennas were employed quasi-simultaneously~(e.g., KTN at 22/43~GHz, and KYS at 86/129~GHz). The polarimetric observables are recorded via the circularly polarized feed horns in the KVN receiver.
        
        The polarimetric properties were measured using eight sets of position-switching scans with a bandwidth of 512~MHz after correcting antenna pointing via cross-scan observation. Several unpolarized sources, mainly planets in the solar system~(Venus, Mars, and Jupiter), were observed to measure and correct polarization leakage terms~(the so-called D-term). We followed the data reduction pipeline introduced in \citet{kang2015} to calibrate the KVN single-dish polarimetric data. The absolute polarization angle of the target sources from the MOGABA was obtained by considering the angle of the Crab nebula in intensity peak position, $\chi_{\rm Crab} = 152\degree$~\citep{aumont2010, weiland2011, planck2016}, at all frequencies~($22-129$~GHz). The quality of the obtained absolute polarization angle in a session was evaluated by comparing the angle of 3C~286 with the literature~\citep[i.e., $33\degree$--$39\degree$, see][]{agudo2012, agudo2018_polami1, perley2013, hull2015, nagai2016}. The absolute polarization angle of the PAGaN data was calibrated using the Crab nebula but at a little bit different reference angle~\citep[for more details, see][]{kam2023}. The difference in reference angles between the two data is less than $3\degree$ at all frequencies, and in most cases, the difference is smaller than the uncertainties in $\chiobs$ of 3C~454.3 at all frequencies. Therefore, the difference in reference angle would not bias the results, so we used all the data together for further analysis.
    
        The conversion of antenna temperature to flux density was performed by scaling the antenna temperature of planets to their brightness temperature. For Jupiter, we referred to the brightness temperatures, $T_{\rm b,22}=136.2\pm0.85$ and $T_{\rm b,43}=155.2\pm0.87$ in units of Kelvin, at 22 and 43~GHz, respectively~\citep{weiland2011}.  We did not use Jupiter for the scaling at 86 and 129~GHz, because of the large angular size compared to the KVN beam sizes and its nonuniform brightness temperature distribution on its surface~\citep{pater2019}. For Mars, we referred to the Mars brightness temperature modeling results at 43, 86, and 129~GHz from the open website\footnote{\url{https://lesia.obspm.fr/perso/emmanuel-lellouch/mars}}. In \citet{dahal2023}, the brightness temperatures of Venus at various radio frequencies were compiled, and we referred to that data.
    
    \subsection{The VLBA 43~GHz}
    \label{sec:obs_vlba}
        About 40 AGN sources have been monitored using the VLBA at 43~GHz by the BU group, and they studied jet kinematics using the monitoring data~\citep{jorstad2017, weaver2022}. Polarimetric characteristics were also investigated by \citet{jorstad2005}. 3C~454.3 is one of the monitoring sources in the program. Thanks to the long baseline length of the VLBA, 3C~454.3 has been resolved into a compact radio core at the end of the east side with an extended jet structure in the northeastern direction. A quasi-stationary component~(Region~C), having a high polarization degree reaching up to a few tens of percent, has been reported. Recently, \citet{traianou2024} studied mas-scale characteristics on 3C~454.3 at 43 and 86~GHz, including polarimetry, between 2013 and 2017. In our work, we utilized the 43-GHz VLBA data in the period from January 2011 to June 2022 to investigate the polarimetric characteristics of the source structures within mas-scale~(such as the core and Region~C) in a more extended period of time. We used the CLEAN algorithm~\citep[e.g.,][]{h"ogbom1974, clark1980, cornwell1999} for imaging and \texttt{modelfit} task with circular Gaussian models in the \texttt{DIFMAP} software~\citep{shepherd1997} to measure the respective locations of the radio core and Region~C.

\section{Results}
\label{sec:result}
    \subsection{Multi-frequency variability}
    \label{sec:lcs}
\begin{figure*}
        \includegraphics[width=0.95\linewidth]{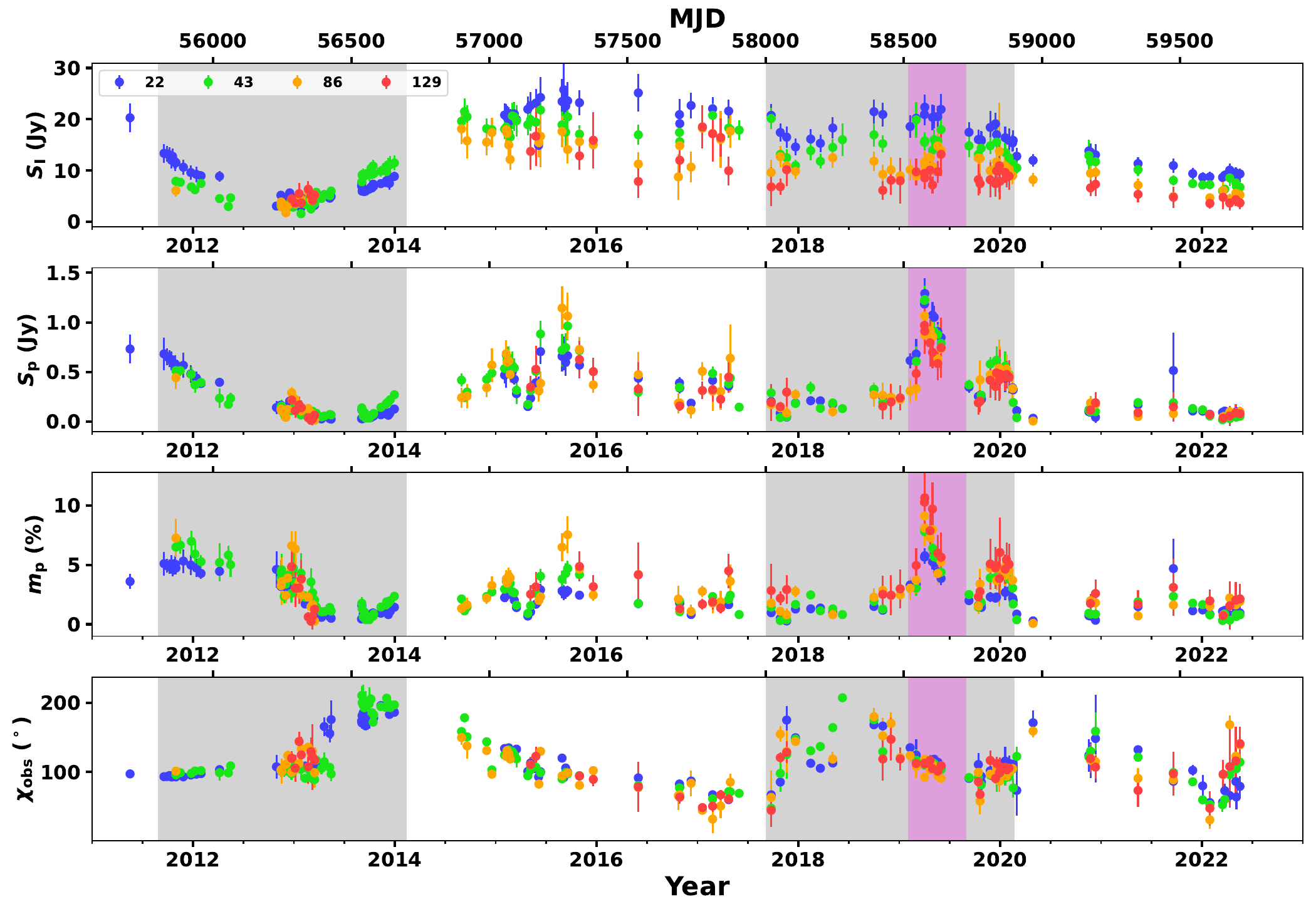}
        \caption{Results of the KVN observations at 22~(blue), 43~(green), 86~(orange), and 129~GHz~(red). Each panel shows total flux density, linearly polarized flux density, polarization degree, and polarization angle from top to bottom. The two time periods that we focus on in this work are shown in the grey-shaded regions and discussed in Section~\ref{sec:lcs}. The magenta-shaded area indicates a flaring period in polarized emission.}
        \label{fig:lc_kvn}
        \end{figure*}

        Figure~\ref{fig:lc_kvn} shows the light curves of 3C~454.3 obtained from the KVN single-dish at 22, 43, 86, and 129~GHz. From top to bottom, each panel shows the variation in the total flux density~($\Si$, Jy), linearly polarized flux density~($\Sp$, mJy), polarization degree~($\Mp$, $\%$), and the observed polarization angle~($\chiobs$, $\degree$). In this study, we resolved $n\pi$-ambiguity of $\chiobs$ in the frequency range of 22--129~GHz in both time and frequency. First, we resolved the $n\pi$-ambiguity of $\chiobs$ in time following \citet{blinov2016}~\citep[see also][]{kiehlmann2016}. In their methodology, they consider significant variation, for instance, more than 90$\degree$ between adjacent epochs as a consequence of $n\pi$-ambiguity. Therefore, if $\Delta\chi_{\rm t} - \sqrt{\sigma_{\chi_{n+1}}^{2} + \sigma_{\chi_{n}}^{2}} > 90\degree$, we resolved $n\pi$-ambiguity in time, where $\Delta\chi_{\rm t}$ and $\sigma_{\chi_{n}}$ are the amount of variation of $\chiobs$ within two adjacent epochs and uncertainty of $n$th measurement, respectively. Then, similarly, we resolved $ n\pi$ ambiguity across all the frequencies in the same epoch by using 22~GHz data as a reference, having better cadence and sensitivity than other frequencies. A similar approach was employed by \citet{agudo2018_polami3} to resolve $n\pi$-ambiguity of 3~mm and 1~mm data, but the authors used an angle difference between the smoothed angle from the previous two data points and the third data point among three data points. The mean cadence of the 22~GHz data in our study is about a month, and the overall percentage of uncertainties in total flux density were $\sim1.8\%$, $\sim1.6\%$, $\sim3.3\%$, and $\sim9.3\%$ at 22, 43, 86, and 129~GHz, respectively. We note that the presented uncertainties are statistical errors. To consider systematic errors from calibration, pointing offset to the source, and opacity correction, we applied additional $10\%$, $10\%$, $15\%$, and $20\%$ of uncertainties at 22, 43, 86, and 129 GHz, respectively\footnote{\citet{lee2016} used $10\%$--$30\%$ of uncertainties in amplitude calibration. However, note that these are from VLBI observations, not single-dish observations.}.
        
\begin{figure*}
        \centering
        \includegraphics[width=0.93\linewidth]{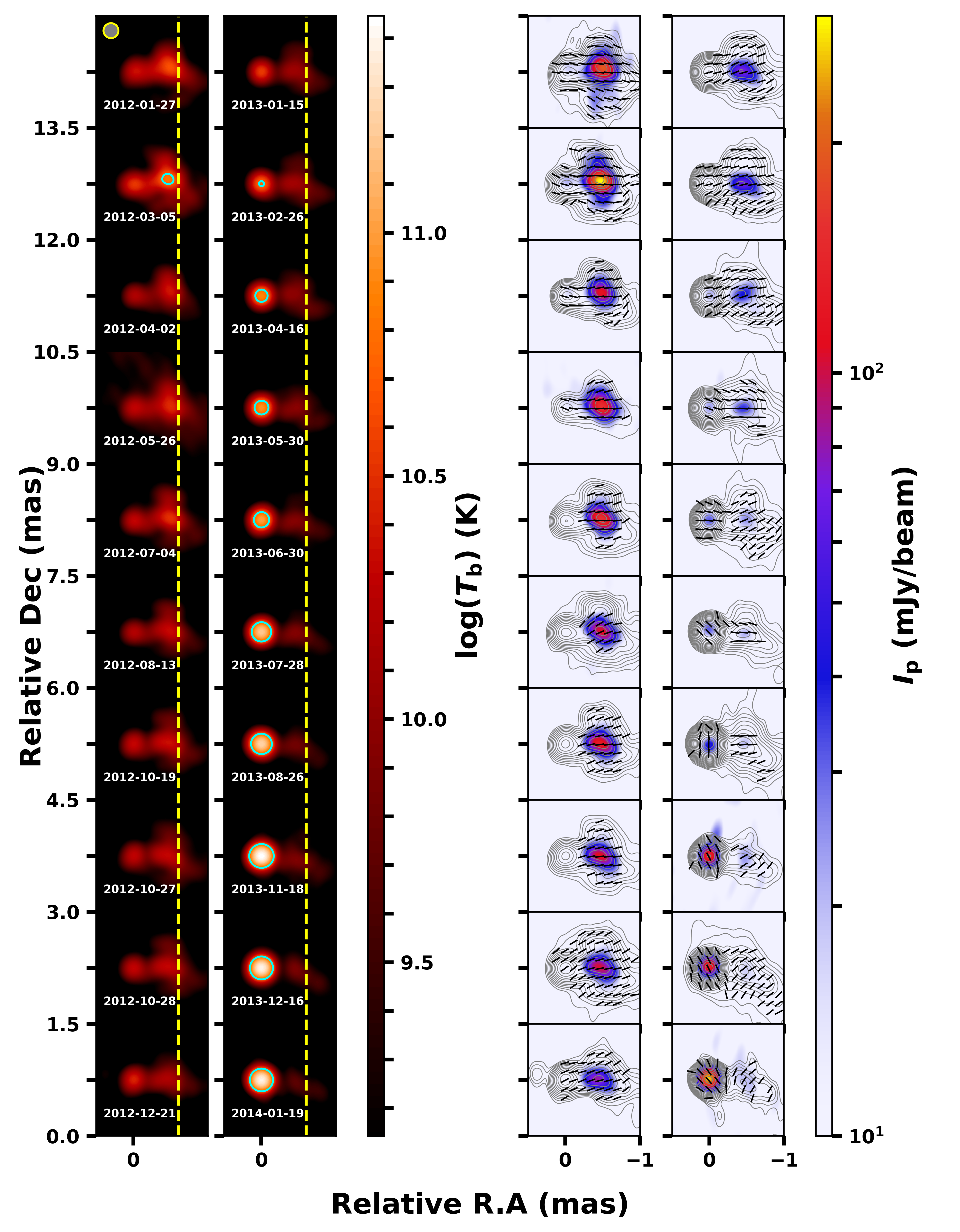}
        \caption{Left$\colon$ The 43-GHz VLBA images in terms of the brightness temperature $T_{\rm b}$ in the source rest frame. The right ascension of the radio core is fixed to 0. The cyan contours indicate where $\Tb=\Teq\approx5\times10^{10}~{\rm K}$. The yellow dashed lines present the location $0.6~{\rm mas}$ away from the core, a typical projected distance of Region~C~\citep[0.45--0.7~mas][]{weaver2022}. Right$\colon$ The contours were obtained from Stokes $I$ images, starting from $3\sigma_{\rm rms}$ level, where $\sigma_{\rm rms}$ represents r.m.s noise level in a residual map. The color maps show linear polarization intensity. All the images are convolved with a beam size of $0.2\times0.2~{\rm mas^{2}}$. The black bars indicate the observed polarization angle on the images where the SNR of the polarized emission is higher than 3.}
        \label{fig:tb_2013}
        \end{figure*}

        In this study, we mainly focus on two periods, around MJD~55800--56700 and MJD~58000--58900. At the beginning of the first period, 3C~454.3 exhibited a global decrease in its flux densities, while both $\Mp$ and $\chiobs$ remained constant until 2013~($\sim5\%$, $\rm \sim 100\degree$). After that, as polarized emission approaches its minimum state around MJD~56400, $\chiobs$ rotates to $\rm \sim 210\degree$, and $\Si$ starts increasing at the same time. To investigate the jet evolution in the polarized emission during this period, we employed the VLBA 43~GHz maps and found different aspects between the core and Region~C, as shown in Figure~\ref{fig:tb_2013}$\colon$the core became brighter while Region~C was fading out. When Region~C dominates the polarized emission, it exhibits $\chiobs\approx90\degree$. However, when the core dominates the emission, $\chiobs$ of the core appears $\sim30\degree$~(i.e., nearly matching $210\degree$ in Figure~\ref{fig:lc_kvn}). These suggest that the evolution in $\Si$ and $\Sp$, and the rotation in $\chiobs$ to $\sim210\degree$ in the KVN data around MJD~56400 is caused by the different evolution of the core and Region~C.
    
        In late 2018~(i.e., second period, around MJD~58400), the angle rotated from $\sim150\degree$ to $\sim100\degree$ at 22, 43 and 86~GHz. At 129~GHz, $\chiobs$ at the beginning point deviates from $150\degree$, but the uncertainty is quite large~($118\angdot8\pm20\angdot9$). This rotation accompanied an increase in $\Sp$, for example, at 129~GHz, by factors of $\gtrsim8$, indicating a polarization flare. The estimated $\Mp$ reached the maximal at all frequencies~(e.g., $\sim10\%$ at 129~GHz as listed in Table~\ref{tab:kvn_pol_result}). We note that the increases in $\Sp$ and $\Mp$ were observed at all four frequencies, and $\Mp$ is clearly larger at higher frequencies around the peak of the flare. The chromatic evolution in $\Mp$ is possibly related to a physical condition, for example, a shocked region~\citep[see Section~\ref{sec:shock-shock}, and see also][]{tavecchio2018}. In addition, the overall trend in variation among polarimetric measurements at all frequencies is comparable. This implies that the flaring event and its polarization variations at 22 to 129~GHz originated from a common emitting region in the relativistic jet.    
    
        To investigate the resolved~(high spatial resolution) polarization characteristics in the jet, the 43-GHz VLBA data was employed. The parameters~($\Ii$, $\Ip$, $\Mp$, and $\chiobs$) were first computed for each pixel in the CLEAN map, then averaged over the pixels associated with each model component. The pixels were chosen from a box area with dimensions matched to the full-width half maximum of the corresponding Gaussian components. The corresponding errors were estimated following \citet{hovatta2012}. The obtained values from the CLEAN flux are consistent with the KVN 43~GHz single-dish data, especially in polarized emission as described in Appendix~\ref{sec:A:appendix_43ghz}.
        
\begin{figure*}
        \includegraphics[width=0.95\linewidth]{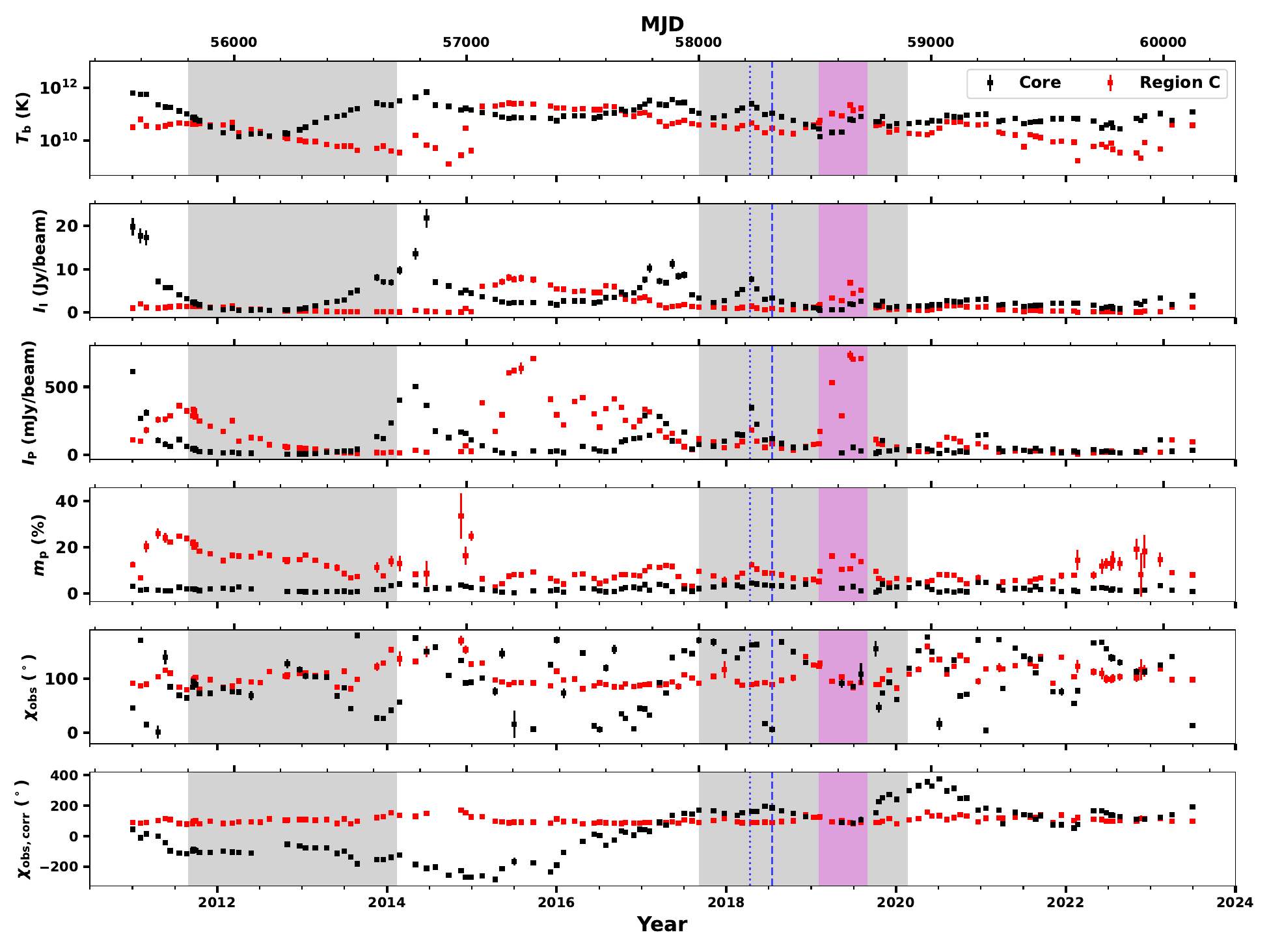}
        \caption{Estimated values of the regions related to the core and Region~C at 43~GHz. Each value was obtained by computing the corresponding pixels. The first four panels indicate brightness temperature, total intensity, linearly polarized intensity, and polarization degree, respectively. The last two panels are polarization angles, but the latter one was obtained after resolving the $n\pi$ ambiguity following the method in \citet{blinov2016}. The blue dotted and dashed lines show the ejection and segmentation time of knot B19~\citep{weaver2022}.}
        \label{fig:lc_bu}
        \end{figure*}

        Figure \ref{fig:lc_bu} shows the light curves of the core~(black dots) and Region~C~(red dots). In general, the core displays higher $\Ii$ than Region~C, while they alternate in dominance in $\Ip$. Several peaks are clearly identified in both $\Ii$ and $\Ip$. 
        
        Although, in general, the core presents relatively low polarization degrees~($\lesssim5\%$, and mean value of $\overline{m}_{\rm p,core}\approx2\%$), the $\chiobs$ of the core suggests rapid variation with $180\degree$ ambiguity to adjacent data points (i.e., in time). After resolving the $n\pi$ ambiguity by assuming the continuous variation described in Section~\ref{sec:lcs}, we obtained a swing (approximately from $-200\degree$ to $+400\degree$) of $\chiobs$ from the core region in a long time scale.
        
        Regarding Region~C, the overall polarization degree $\Mp$ is higher than at the core, with the maximum value reaching $\sim20\%$~(and mean value of $\overline{m}_{\rm p,C}\approx10\%$). The large uncertainties on $\Mp$ arise when the signal-to-noise ratio of Region~C is low. Interestingly, $\chiobs$ of this region is aligned to be $\sim100\degree$ in a high-polarization state~(e.g., $\gtrsim$200~mJy/beam) and deviates from that in low-polarization~(e.g., <200~mJy/beam) periods. In addition, several clear peaks in the polarized emission of Region~C have seemed to appear after the flaring activities in the core. Given that Region~C is the downstream region beyond the core, it is plausible to interpret that a moving knot induces activity when the knot arrives at Region~C. \citet{jeong2023} suggested that the $\gamma$-ray flare in August 2015~(about MJD~57250) is attributed to knot K14~\citep[B12 knot defined in][]{weaver2022} when it arrived at Region~C.
        
\begin{figure*}
        \centering
        \includegraphics[width=1\linewidth]{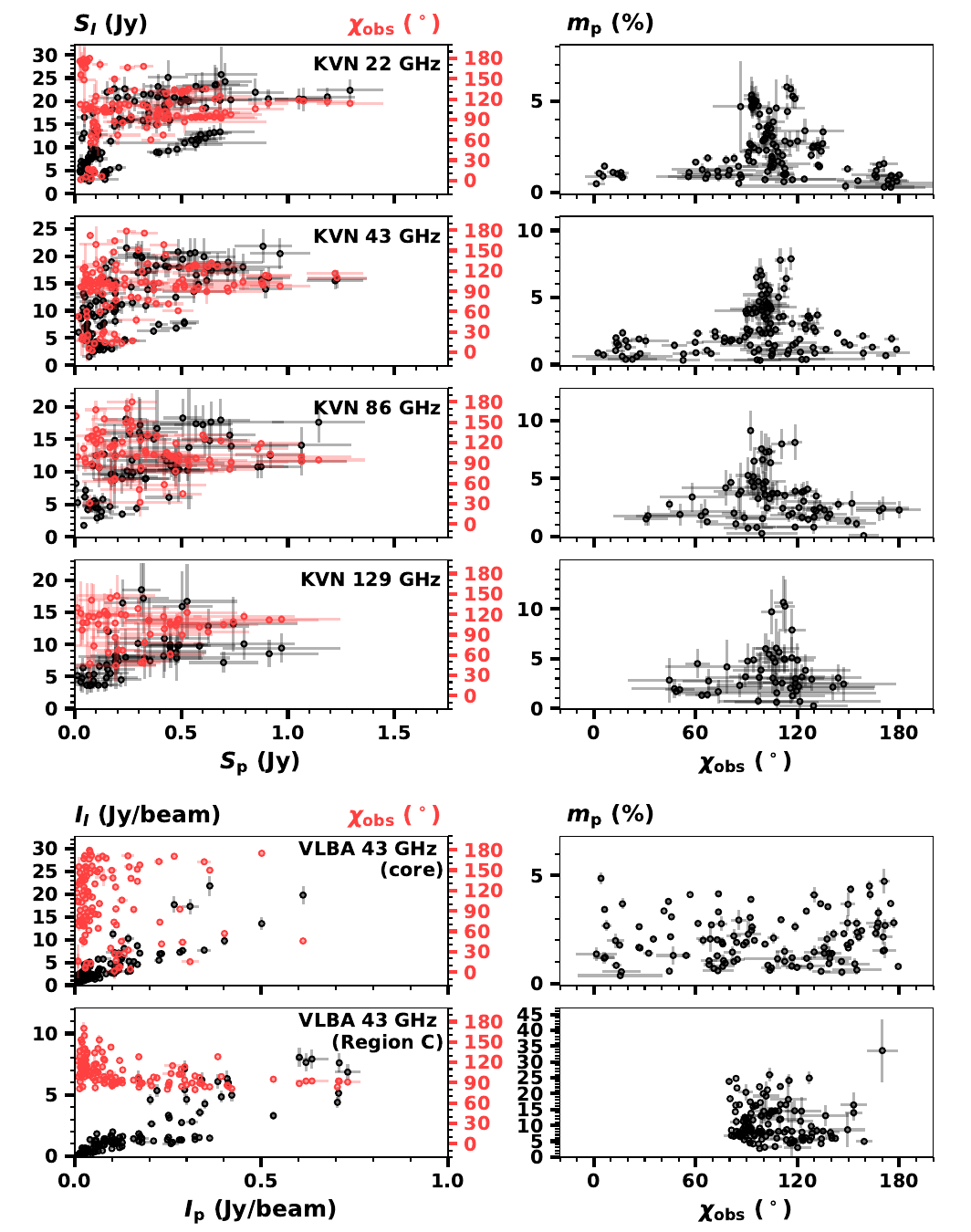}
        \caption{Left$\colon$ Dependence of total emission~(black symbols) and $\chiobs$~(red symbols) on polarized emission for the KVN data at different frequencies~(the first 4 panels) and the two different jet regions from the VLBA 43~GHz, the core and Region C~(the last 2 panels). Right$\colon$ Dependence of $\Mp$ on $\chiobs$.}
        \label{fig:pol_correlation}
        \end{figure*}

        Correlations between the polarimetric measurements obtained from the KVN single-dish and the VLBA were investigated, as shown in Figure~\ref{fig:pol_correlation}$\colon\Si/\chiobs$--$\Sp$ and $\Ii/\chiobs$--$\Ip$~(left) and $\Mp$--$\chiobs$~(right). $\chiobs$ measured from the KVN were distributed with respect to $\sim100\degree$ at all four frequencies when $\Sp$ was bright. Similar behavior was also found in Region~C. The right panel also confirms such behavior of $\chiobs$. These indicate that a physical process increasing the polarized emission in Region~C causes the observed relations.
        
    \subsection{Polarization angle and rotation measure}
    \label{sec:pa_rm}
\begin{figure*}
        \includegraphics[width=0.95\linewidth]{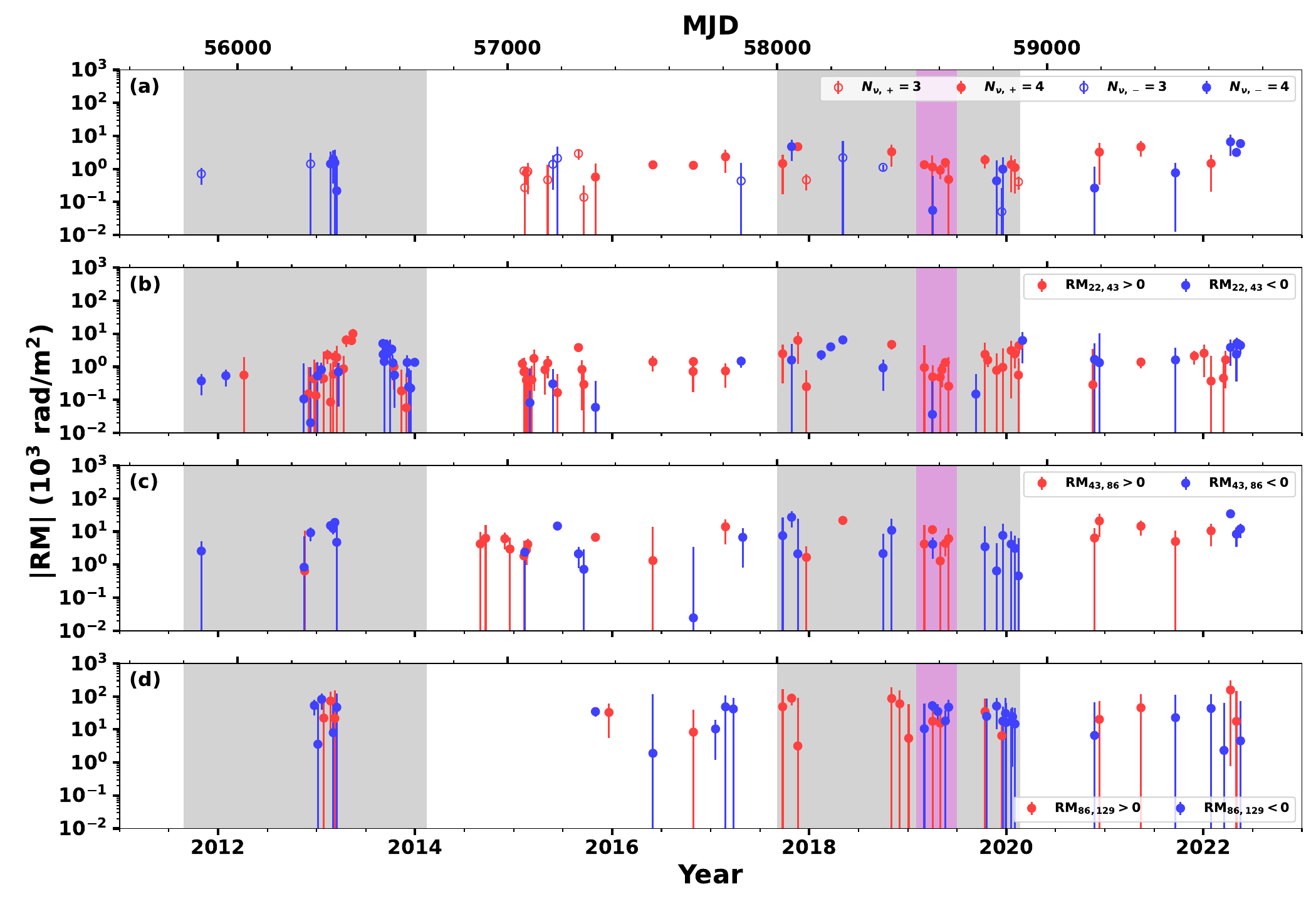}
        \caption{RMs in the first panel~(panel a) were obtained via linear fit~($\chi=\chi_{0}+{\rm RM}\,\lambda^{2}$) if more than two data points were available. There are several large uncertainties~($\sigma_{\rm RM}\gtrsim{\absrm}$) attributed mainly to a significant deviation in $\chiobs$ to a linear relation. From the second to the last panels, RMs were calculated in pairs of 22--43~GHz~(panel b), 43--86~GHz~(panel c), and 86--129~GHz~(panel d). RM signs are shown in color$\colon$blue and red for negative and positive signs, respectively. The filled and open circles in the first panel indicate the number of frequencies used in the fitting, four and three data points, respectively. Each shaded area is the same as in Figure~\ref{fig:lc_kvn}.}
        \label{fig:rm_pairs}
        \end{figure*}

        In this section, we present the results of RM using the KVN single-dish data. RM is a measurement that shows the Faraday screening effect of a foreground medium. As noted in Equation \ref{eq:RM}~(${\rm RM} \propto \int\Ne B_{\parallel}dl$), high RM indicates high $\Ne$, strong $B_{\parallel}$, or both. Figure~\ref{fig:rm_pairs}a presents the RMs obtained from a linear model fitting. The RMs are shown in a log scale to illustrate the variation in magnitude, while the RM signs are described by red dots for a positive RM~(i.e., parallel magnetic field toward us) and in blue dots for a negative RM~(i.e., magnetic field in the opposite direction). The number of frequencies in the linear fitting is described by filled and open circles, respectively.
    
        We found that the order of RM values obtained from the linear fitting lie in a level of $\sim10^{3}~{\rm rad\,m^{-2}}$ and the mean value is approximately $(-0.1\pm2.7)\times10^{2}~{\rm rad\,m^{-2}}$. Several sign flips appear, but they are not significant, given their uncertainty ranges. In a previous polarization monitoring study, the averaged RM of 3C~454.3 was reported to be $(-2\pm52)\times10^{3}~{\rm rad\,m^{-2}}$ at 86--229~GHz~\citep{agudo2018_polami3}. Similarly, the absolute values in RM~($\absrm$) suggest a variability in $\absrm$ as well. In addition, in the fourth column in Figure~B.1\footnote{Appendix B Plots are available on \url{https://zenodo.org/records/14967589}}, multiple epochs show $\chiobs$ values with large offsets from a single RM fit.
        
        In order to focus on the epochs during which a single component dominates the polarized emission, the VLBA data was employed. In this work, we studied the periods when Region~C dominates the polarized emission. Based on $\Mp$ obtained from the KVN and 43-GHz VLBA images~(Figure~\ref{fig:tb_2019flare}), several epochs were selected as C-dominating periods$\colon$ from February 2019 to June 2019. During these periods, $\Mp$ is high with chromatic behavior in frequency, indicating partially ordered magnetic field lines~(see Figure~B.1). In particular, we note that the amount of increased $\Sp$ is consistent at all frequencies and is remarkable compared to the pre-flare state. These results tentatively suggest that the polarimetric variations are mainly related to the evolution of Region~C. The obtained $\chiobs$ at all frequencies are aligned near $100\degree$, but it does not follow a simple linear relation with $\lambda^{2}$.

\section{Discussion}
\label{sec:discussion}
    \subsection{Total and polarized brightness evolution}
    \label{sec:disc_flux-evolution}
        Figure \ref{fig:lc_kvn} shows the time variation in flux densities~($\Si$ and $\Sp$), polarization degree~($\Mp)$, and polarization angle~($\chiobs$) obtained from the KVN multi-frequency single-dish observations. In this Section, we discuss the variations in two periods, MJD~55800--56700 and MJD~58000--58900, shaded with grey color.
        
        In the period of MJD~55800--56100, both $\Si$ and $\Sp$ decreased by a factor of about $0.6$, while $\chiobs$ remained nearly constant~(i.e., $\sim100\degree$) at 22 and 43~GHz. However, starting around MJD~56300, the light curves display different evolution between total and polarized emission with rotation in $\chiobs$ from $\sim100\degree$ to $\sim200\degree\colon$ increase in total flux density with a decrease in polarized flux density. The different evolution is unlikely to originate from a single synchrotron emission region. If there were two or more emitting regions observed in a single-dish beam, the different evolution in $\Si$ and $\Sp$ can be explained$\colon$ for example, one region dominant in polarized emission has diminished~(decreasing $\Sp$), and then the other weakly polarized and fainter one becomes brighter and dominates the overall emission. Indeed, it seems that two main regions dominate the overall emission in 3C~454.3 around this period$\colon$ the core and the quasi-stationary Region~C. As shown in Figure \ref{fig:lc_bu}, while the core was bright in total intensity, the polarized intensity was predominant in Region~C at the beginning of the first period. After that, Region~C decayed in polarized emission with a nearly constant $\chiobs$ while the core started increasing its brightness around MJD~56300.
        
        In the beginning of the second period~(MJD~58000--58900), while the source was bright in $\Si$ it was not in $\Sp$~(Figure~\ref{fig:lc_kvn}). However, starting from MJD~58400, both total and polarized emission increased at all the KVN frequencies, reaching $\Sp\sim1000~{\rm mJy}$, and $\Mp\sim10\%$ in April 2019. The obtained $\Mp$ exhibits chromatic behavior during the increasing stage. The amount of increase in polarized emission is similar among the frequencies. Unlike the behavior presented in the first period, the increase in both total and polarized emission may indicate a single highly polarized synchrotron emitting region leading the polarization evolution. In addition, we found that such activity also appears in Region~C with $\Ip > {\rm 500~mJy/beam}$, as shown in Figure~\ref{fig:lc_bu}; thus, in polarized emission, Region~C was dominant and the core was quiet. Combining these results, we conclude that Region~C is the origin of the polarization flare in 2019 at frequencies between 22 and 129 GHz.
    
    \subsection{The origin of variation in polarization angle}
    \label{sec:disc_pa_variation}
        In this study, we applied the methodology introduced in \citet{blinov2016} for resolving $n\pi$ ambiguity in time as described in Section~\ref{sec:lcs}. As a cross-check, we also determined that the $\chiobs$ obtained from the VLBA data agree close to that of the KVN at 43~GHz after resolving the ambiguity~(Figure~\ref{fig:lc_kvn-bu}).
        
        The resultant angles obtained from the KVN show several variations~(Figure~\ref{fig:lc_kvn}). Even with such a variation, the polarimetric relations of $\chiobs$ reveal angles aligned to $\sim 100\degree$ when strongly polarized at all frequencies~(Figure~\ref{fig:pol_correlation}). This behavior is attributed to Region~C as described in Section~\ref{sec:lcs}, indicating that, in general, polarized emission is dominated by Region~C when the source is strongly polarized.
        
        Based on the VLBA 43~GHz data, we separately investigated the $\chiobs$ of the core and Region~C using the corresponding pixels where each region is located, as shown in Figure~\ref{fig:lc_bu}. Interestingly, the core seems to suggest a long-term variation in $\chiobs$ after resolving a time-$n\pi$ ambiguity. Possible explanations for this long-term variation are (i)~helical magnetic field geometry with a bending within the core~\citep[for this case, one may find in][]{myserlis2018}, (ii)~rotation or $\sim90\degree$ of flip in $\chiobs$ by an opacity effect~\citep{lobanov1998, gabuzda2001, porth2011, chamani2023}, or (iii)~by a stochastic random walk process~\citep{blinov2015, kiehlmann2017}. Alternatively, it is also possible that the variation in $\chiobs$ is a biased result due to a relatively large cadence~(about a month). Since the method we applied assumes no significant jump in $\chiobs$~($>90\degree$) between the adjacent epochs, the related information may be biased if the intrinsic angle rotates faster than the observed cadence. For example, a rapid variation~(e.g., spanning a few hours to a few days) in $\chiobs$ can occur via magnetic reconnection~\citep{zhang2018, zhang2020}. Possibly, high-cadence and long-term VLBI observations in the future would reveal whether this long-term variation is intrinsic or not.
        
        Meanwhile, Region~C suggests a specific angle, mostly between $90\degree$--$120\degree$. Among the observations collected in this study, one of the worthy periods to note is around April 2019, showing a prominent polarization flaring activity in Region~C, as identified in Section~\ref{sec:disc_flux-evolution}~(see also Figure~\ref{fig:lc_bu}). As the flare grows up from the beginning~($\rm \sim MJD~58400$), $\chiobs$ values at all frequencies rotate into a direction parallel with the jet axis ~\citep[$\sim-80\degree$,][]{weaver2022}. Such a rotation in $\chiobs$ with increased flux density implies a physical process in Region~C. A polarization angle parallel to the jet axis is feasible through a dominant toroidal magnetic field component in the jet.
    
    \subsection{Faraday rotation measure in arcsecond scales}
    \label{sec:disc_kvn_rm}
        The polarimetric measurements~($\chiobs$ and $\rm RM$) used in Figure~B.1 are listed in Table~\ref{tab:kvn_pol_result} and Table~\ref{tab:kvn_rm_result}. Figure~\ref{fig:rm_pairs}b--d and B.2 show $\absrm$ obtained from the KVN in the frequency domain. As presented, RM shows a large mean of $\absrm \gtrsim 10^{3} {\rm rad/m^2}$ as well as a variability not only in its magnitude but also in its sign. These imply that an external medium is unlikely to be the main reason for the Faraday screening medium if present.

        The calculated mean of $\absrm$ in the available periods are $2.9\pm0.2$, $7.4\pm1.3$, and $32.6\pm11.6$ in units of $\rm 10^{3}rad\,m^{-2}$ at 22--43, 43--86, and 86--129 GHz pairs, respectively~(see Section~\ref{sec:pa_rm}). Although it seems to suggest frequency-dependent $\absrm$, one needs to be careful in interpreting frequency dependency in $\absrm$ as systematic effects may affect $\absrm$.

        There are several origins of the time variability and frequency dependence of RM. For example, the variability can be attributed to dynamics in the jet, such as turbulence or helical magnetic field lines, since it depends on the direction of magnetic field lines toward us. \citet{zamaninasab2013} found evidence of a large-scale helical field structure in the jet of 3C~454.3 through multi-frequency VLBI observations.
        
        Alternatively, we could not rule out the possible beam convolution effect (i.e., beam depolarization) on RM variability at other frequencies, although we have identified consistent polarized emission of the source in arcsecond and mas scales at 43 GHz. The source has a complex jet structure~(in mas-scale) not only in total emission but also in polarized emission, as shown in Figure~\ref{fig:tb_2013} and \ref{fig:tb_2019flare}. The complex jet structure with large beam sizes in the KVN single-dish observations possibly results in rotation in $\chiobs$ through beam convolution effect attributing to $\absrm$ variation. For this reason, we selected a few epochs~(from February 2019 to June 2019) in which Region C dominates the polarized emission of 3C~454.3 at 43~GHz and found the frequency dependence in $\absrm$ in those epochs.
        
        To understand the frequency-dependence $\absrm$ of 3C~454.3 shown in this work in more detail, we consider the opacity effect to $\chiobs$. As the synchrotron emission becomes optically thick at a low frequency, it is possible that the opacity effect rotates $\chiobs$ about $90\degree$~\citep{gabuzda2001, porth2011}. Alternatively, the frequency dependence can be related to the core-shift effect~\citep{lobanov1998}. As described in \citet{jorstad2007}, if the Faraday screening medium is very close to the jet, such as a sheath layer, frequency-dependent $\absrm$ can be obtained.
        
        However, the downstream jet, Region C, is expected to be optically thin in the KVN observing frequency range, restricting us from considering the opacity effect as the origin of the frequency-dependent $\absrm$. In addition, we cannot be confident if the polarized emission from the core is faint at higher frequencies~(86 and 129~GHz). If the core has prominent polarized emission compared to Region C at these frequencies, then $\chiobs$ may be affected by this.
        
        We consider that the frequency-dependent $\absrm$ shown in this work is most likely attributed to complex and unresolved polarized emission regions within 1~mas~(i.e., the case of beam convolution) rather than an external Faraday screening medium. \citet{traianou2024} studied spatially resolved polarized emission in 3C~454.3 using VLBI observations at 43 and 86~GHz. As shown in their work, the core and Region C have different spectral indices at 43 and 86~GHz$\colon$ flat and thin spectra for the core and Region C, respectively, resulting in different morphologies in total intensity and polarization at 43 and 86~GHz~(i.e., showing complex structure). Therefore, high-resolution multi-frequency VLBI observations in full-polarization mode would be required to investigate a more detailed analysis of RM variability in the jet.
    
    \subsection{Origins of a polarization flare in Region~C}
    \label{sec:disc_origin_of_C}
        In Section~\ref{sec:disc_flux-evolution}, we identified that the flaring region is related to Region~C based on a consistent evolution in flux density across the KVN frequencies, and based on the VLBA 43~GHz maps. In addition, $\chiobs$ rotates from $\sim150\degree$ to $\sim100\degree$ during the flare. Since the core region was faint with several non-detections in polarization during this period, we interpret that the increasing $\absrm$ with frequency and other polarimetric properties are related to a physical process in Region~C. To understand it, we consider three cases$\colon$ (i) jet-medium interaction, (ii) magnetic reconnection, and (iii) shock--shock interaction. Finally, we investigate how the Doppler boosting effect via a change in viewing angle can describe the observed brightness increase.
        
        \subsubsection{Jet-medium interaction}
        \label{sec:jmcollision}
            For the first scenario, a jet interacting with an ambient medium may result in a high $\rm RM$. Broad-line region~(BLR) and narrow-line region~(NLR) are the possible candidates for such a medium. In addition, the jet can be bent toward us by the collision, leading to an increase in the Doppler factor and, thus, the brightness as well. However, we note the de-projected location of Region~C is $d_{\rm C}\approx200~{\rm pc}$ away from the core by adopting a mean viewing angle $\theta_{\rm v,mean}\approx1\angdot4$~\citep[see Table 12 in][]{weaver2022}, and its projected distance from the core of $\sim0.6~{\rm mas}$. The distance of BLR in 3C~454.3 is $d_{\rm BLR}\approx0.2~{\rm pc}$~\citep{bonnoli2011, costamante2018}, which was estimated using the size-luminosity relation. Moreover, based on the activity in optical line emission, \citet{amaya2021} interpreted that the increase in continuum emission during a flaring period~(FP2, approximately 2014--2017), in which Region~C brightened, was not strongly related to the BLR. Therefore, we exclude the BLR from the dense cloud candidate. If the medium corresponds to the NLR at scales of pc to kpc, we expect a stable $\Ne$, $B_{\parallel}$, and the path length. In this case, the estimated RM magnitude and its sign would not vary significantly either in time or frequency. However, we would expect constant $\rm RM$ across the observing frequencies and time if the Faraday screening medium is related to the NLR. Therefore, we suggest that the NLR is also less likely to cause the flare with variation in $\rm RM$ via jet-medium collision.
    
        \subsubsection{Magnetic reconnection}
        \label{sec:mreconnect}
            Alternatively, magnetic reconnection, a physical process of rejoining two oppositely aligned magnetic field lines, can introduce increased $\Ne$, magnetic field strength, and the acceleration of plasmoids in the jet region~\citep{zhang2018, nathanail2020, sironi2021}. Given the magnetic energy density dominance in Region~C, as reported in \citet{jeong2023}, this region is suitable for magnetic reconnection. The magnetic dominance can also be supported by having a brightness temperature~($\Tb$) lower than the equipartition condition~\citep[$\Teq\approx5\times10^{10}~{\rm K}$,][]{readhead1994}. Figure~\ref{fig:tb_2019flare} shows the $\Tb$ estimated in the source rest frame. The obtained $\Tb$ of Region~C was lower than $\Teq$ before a moving knot arrived in Region~C, indicating a possible magnetic dominance.
            
            By using particle-in-cell~(PIC) simulation, \citet{petropoulou2016} studied the physics of plasmoids in a relativistic jet, which is formed by the fragmentation of reconnection current sheets~(current sheets derived by magnetohydrodynamic instability) which is a region containing magnetic field lines and energetic particles. From the simulation, they found that the highest emission is observed when a plasmoid leaves the current sheet layer. \citet{zhang2020} investigated the signatures in the polarization of magnetic reconnection in blazar at multi-wavelength. In their work, magnetic field lines are found to be disordered when a reconnection layer segments into several plasmoids. This results in a low $\Mp$ when the observed total emission peaks due to depolarization by the tangled magnetic field lines in a plasmoid. In our study, however, we obtained increasing $\Mp$ and $\Sp$ as $\Si$ increases during the flare in 2019. Therefore, we consider that magnetic reconnection is less likely to govern the physics in Region~C.
    
        \subsubsection{Shock--shock interaction}
        \label{sec:shock-shock}
            In the last scenario, $\Ne$ and $B_{\parallel}$ in a jet can also be increased via shock--shock interaction, leading to compression of a jet region. Previous studies on 3C~454.3 at high resolution have identified a quasi-stationary behavior of Region~C, although a marginal motion was observed at a distance range of 0.45--0.7~${\rm mas}$ from the core~\citep{jorstad2017, weaver2022}. The motion is possibly related to the intrinsic physics of Region~C, but as suggested in \citet{weaver2022}, it also can be explained by the relative motion of the core produced by the core-shift effect. \citet{liodakis2020} explained the spatial variation in Region~C with the emergence of a new jet component from the core. 
            
            The possibility of Region C as a recollimation or standing shock has been discussed in previous studies~\citep{gomez1999, jeong2023, traianou2024}. If Region~C is related to a shock, its interaction with an upstream shock moving downstream is considered a shock--shock interaction. In this case, the intrinsic polarization angle of the shocked region will be parallel with the jet axis due to the compression of magnetic field components perpendicular to the jet axis~\citep{marscher2014} for the case of conical shock or self-generated magnetic field lines that are predominantly aligned in a direction parallel to the shock front~\citep{tavecchio2018}. 
            
            Moreover, in the shock--shock interaction scenario, a chromatic behavior in $\Mp$ with observing frequency is expected based on the energy-stratified model~\citep{tavecchio2018, tavecchio2020, liodakis2022}. In the region where the jet is most compressed, such as the Mach disk or shock front, the magnetic field lines are expected to be ordered, and thus a high $\Mp$ is observed. Beyond the region~(downstream the jet), the jet expands, and the magnetic field lines are relatively disordered as it flows downstream, leading to low $\Mp$. Since high-energy electrons undergo a stronger radiative cooling process than low-energy electrons, the dominant emission region at higher frequencies is closer to the compressing region than at lower frequencies. As shown in Figure~B.1 and Table~\ref{tab:kvn_pol_result}, $\Mp$ is as high as $\sim10\%$ at high frequencies but is rather low~($\sim5\%$) at 22 and 43~GHz during the flare. If the extended jet~(e.g., $\gtrsim1$~mas, see also Figure~\ref{fig:tb_2019flare}) of 3C~454.3 is highly polarized, $\Mp$ can be depolarized via the beam depolarization effect, especially at a low frequency. However, $\Sp$ and $\Mp$ are comparable between scales at mas and arcsecond, as shown in Figure~\ref{fig:lc_kvn-bu}, and thus, we expect the extended jet to have a negligible effect on the polarized emission at frequencies higher than 43~GHz. Furthermore, we also expect a negligible depolarization effect by turbulent magnetic fields if Region~C is related to the shock interaction showing well-ordered magnetic field lines.

            It is also possible that Region C is related to an oblique shock rather than a simple recollimation or standing shock. In the case of an oblique shock, jet direction potentially varies by jet instabilities or jet bending. In particular, the jet of 3C~454.3 has a very narrow viewing angle~($\vang$), so its brightness is sensitive to jet flow direction via the Doppler boosting effect as discussed in a previous study~\citep{traianou2024}. Therefore, it is essential to examine if the pure jet bending can explain the observed brightness increase during the polarization flare.
    
        \subsubsection{Changes in viewing angle}
        \label{sec:vangle}
            As blazar jets generally have $\vang$ smaller than $10\degree$, a change in $\vang$ can induce strong variability. \citet{traianou2024} suggested Region C as a bent jet region to explain the disappearance of a moving knot and brightness increase. In a similar way, we investigated the variation in Doppler factor $\delta$ to determine whether the increased $\Si$ and $\Sp$ during the flare were strongly attributed to the change in $\vang$. As shown in Figure~\ref{fig:tb_2019flare}, a plasma blob having $\Tb\gtrsim\Teq$ was identified and ejected downstream from the core to Region~C. Given its ejection time, the plasma blob is likely related to the knot B19 defined in \citet{weaver2022}. As the blob approaches Region~C, $\Tb$ increases with a peak value of $\sim2.6\times10^{11}~{\rm K}$. Compared to the minimum peak value of the blob~($T_{\rm b,min}\approx7.3\times10^{10}~{\rm K}$) during its ejecting period, the brightness increased by a factor of more than three. After crossing Region~C, the blob seems to move further downstream with reduced $\Tb$. We note that $\Tb$ is computed in the source rest frame, so the only factors that can change $\Tb$ are particle acceleration/energy loss or a change in viewing angle that affects $\delta$.
            
            To evaluate the Doppler boosting effect by viewing angle, we adopt the apparent speed of knot B19 and the mean viewing angle of 3C~454.3 as $\beta_{\rm app}=18\pm1.8$ and $\theta_{\rm v,mean}=1\angdot4\pm0\angdot4$, respectively~\citep[see Table 8 and 12 in][]{weaver2022}. At this angle, we obtain the bulk Lorentz factor of B19 as $\Gamma\approx21.7\pm1.4$. By assuming a constant $\Gamma$ of the knot during its journey, we find the ratio of the Doppler factors at viewing angles of $\vang=1\angdot4\pm0\angdot4$ and $0\degree$ to be $\sim1.28\pm0.16$. Based on the observed ratio of $\delta$ larger than three, we conclude that by itself, a change in $\vang$ is insufficient to describe the increased $\Tb$, implying an additional source intrinsic process, such as particle acceleration.
            
            Although the observational results favor an additional particle acceleration in Region~C by a shock--shock interaction, the lack of emission beyond the region leaves an open question. If the region has a very steep optically thin spectrum, for example, due to a strong cooling process, this can explain the absence of emission. A high-resolution analysis at multi-frequency, including low-frequency~(e.g., 15~GHz), would be required to investigate spectral information beyond Region~C.

\section{Conclusions}
\label{sec:conclusions}
    In this work, we studied the polarimetric variations of 3C~454.3 using decadal~(2011--2022) data at millimeter wavelengths. We found several notable polarimetric features, especially in 2013--2014, and 2019. In this Section, we summarize our results and conclusions as follows$\colon$
    
    \begin{enumerate}
        \item At all frequencies used in this work, relations between the polarimetric characteristics show a trend of $\chiobs\approx100\degree$ when the polarized emission is strong~(Figure~\ref{fig:pol_correlation}). By analyzing the VLBA data, we found that Region~C yields consistent results, indicating a prominent polarized jet region downstream of the radio core even at 129~GHz. While the $\chiobs$ of Region~C does not change after resolving the $n\pi$ ambiguity in time~\citep{blinov2016}, the $\chiobs$ of the radio core exhibits a large long-term swing.
    
        \item Around 2013--2014, $\chiobs$ rotated from $\sim100\degree$ to $\sim200\degree$ with decreasing $\Sp$ and prominent variations in $\absrm$ at 22--43~GHz pair. From the VLBA 43~GHz data, we found that Region~C dominated the polarized emission in mas-scale and then decayed in all emission~($\Si$ and $\Sp$) during this period. At the same time, the core emission was enhanced. These imply that the rotation in $\chiobs$ and the $\absrm$ variation did not originate from a common emitting region but can be attributed to the co-evolution of Region~C and the core.
    
        \item A notable polarization flaring event in 2019 was captured in the KVN observations. Even though this activity increased the total emission a bit, a prominent evolution appeared in the polarized emission. During the flaring period, the $\chiobs$ at all the KVN observing frequencies rotated toward $\sim100\degree$ with partially ordered magnetic field lines~(e.g., $\Mp\gtrsim10\%$ at 129~GHz). By investigating high angular resolution maps from the VLBA at 43~GHz, we identified that the polarization flare occurred in Region~C when a moving blob~\citep[possibly related to knot B19 in][]{weaver2022} arrived at Region~C.
    
        \item To describe the flare in Region~C and to investigate which physical process is favorable for the flare, we considered several cases. (i) Jet-cloud interaction via BLR or NLR is unlikely to cause the flare based on the de-projected distance of Region~C, the lack of optical line emission~\citep{amaya2021}, and variable RM. (ii) Magnetic reconnection is excluded due to the opposite evolution in $\Mp$ between the PIC simulation~\citep[decreasing $\Mp$,][]{zhang2020} and our observations~(increasing $\Mp$) during the flare. (iii) Shock--shock interaction can explain the overall observational results -- chromatic $\Mp$, concurrent rise in $\Si$ and $\Mp$, and $\chiobs$ parallel to the jet axis~\citep{marscher2014, tavecchio2018, tavecchio2020, liodakis2020}. Therefore, we suggest that a shock--shock interaction is the most favored mechanism for the flare in 2019.
    
        \item We also investigated a possible beaming effect caused by a local bending of the jet for some reason. The mean $\vang$ of 3C~454.3 was obtained as $1\angdot4\pm0\angdot4$ by averaging those of several knots~\citep{weaver2022}. By adopting the apparent speed of knot B19 as $\beta_{\rm app}\approx18\pm1.8$, we obtained the resultant Doppler factor $\delta\approx33.9\pm4.4$ for the knot. In the assumption of constant $\Gamma$ of the moving knot, the Doppler factor at $\vang=0\degree$ was computed to be $\delta_{0}\approx43.5\pm2.8$. Therefore, the change in $\vang$ toward us can increase $\delta$ by $1.28\pm0.16$, and this is insufficient to explain the observed increased brightness of about $3.5$ times. Provided with this and the favored physical model, we interpret that the shock--shock interaction may be responsible for the polarization flaring activity in Region~C with additional particle acceleration.
    \end{enumerate}

\section*{Data Availiability}
The Appendix B Figures are available on \url{https://zenodo.org/records/14967589}.

\begin{acknowledgements}
    We thank Thomas Krichbaum, Alan Marscher, and Svetlana Jorstad for reading the manuscript and helping us to improve it with valuable comments. The KVN is a facility operated by the KASI~(Korea Astronomy and Space Science Institute). The KVN observations and correlations are supported through high-speed network connections among the KVN sites provided by the KREONET~(Korea Research Environment Open NETwork), which is managed and operated by KISTI~(Korea Institute of Science and Technology Information). This work was supported by the National Research Foundation of Korea (NRF) grant funded by the Korea government (MIST) (2020R1A2C2009003). We also acknowledge financial support from the National Research Foundation of Korea (NRF) grant 2022R1F1A1075115. This study makes use of VLBA data from the VLBA-BU Blazar Monitoring Program~(BEAM-ME and VLBA-BU-BLAZAR; http://www.bu.edu/blazars/BEAM-ME.html), funded by NASA through the Fermi Guest Investigator Program. The VLBA is an instrument of the National Radio Astronomy Observatory. The National Radio Astronomy Observatory is a facility of the National Science Foundation operated by Associated Universities, Inc. This work made use of Astropy:\footnote{http://www.astropy.org} a community-developed core Python package and an ecosystem of tools and resources for astronomy~\citep{astropy:2013, astropy:2018, astropy:2022}.

\end{acknowledgements}

\bibliographystyle{aa} 
\bibliography{main} 

\begin{appendix}
\onecolumn
\section{Comparison of emission at scales of mas and arcsecond}
\label{sec:A:appendix_43ghz}
\begin{figure}[H]
    \includegraphics[width=0.95\linewidth]{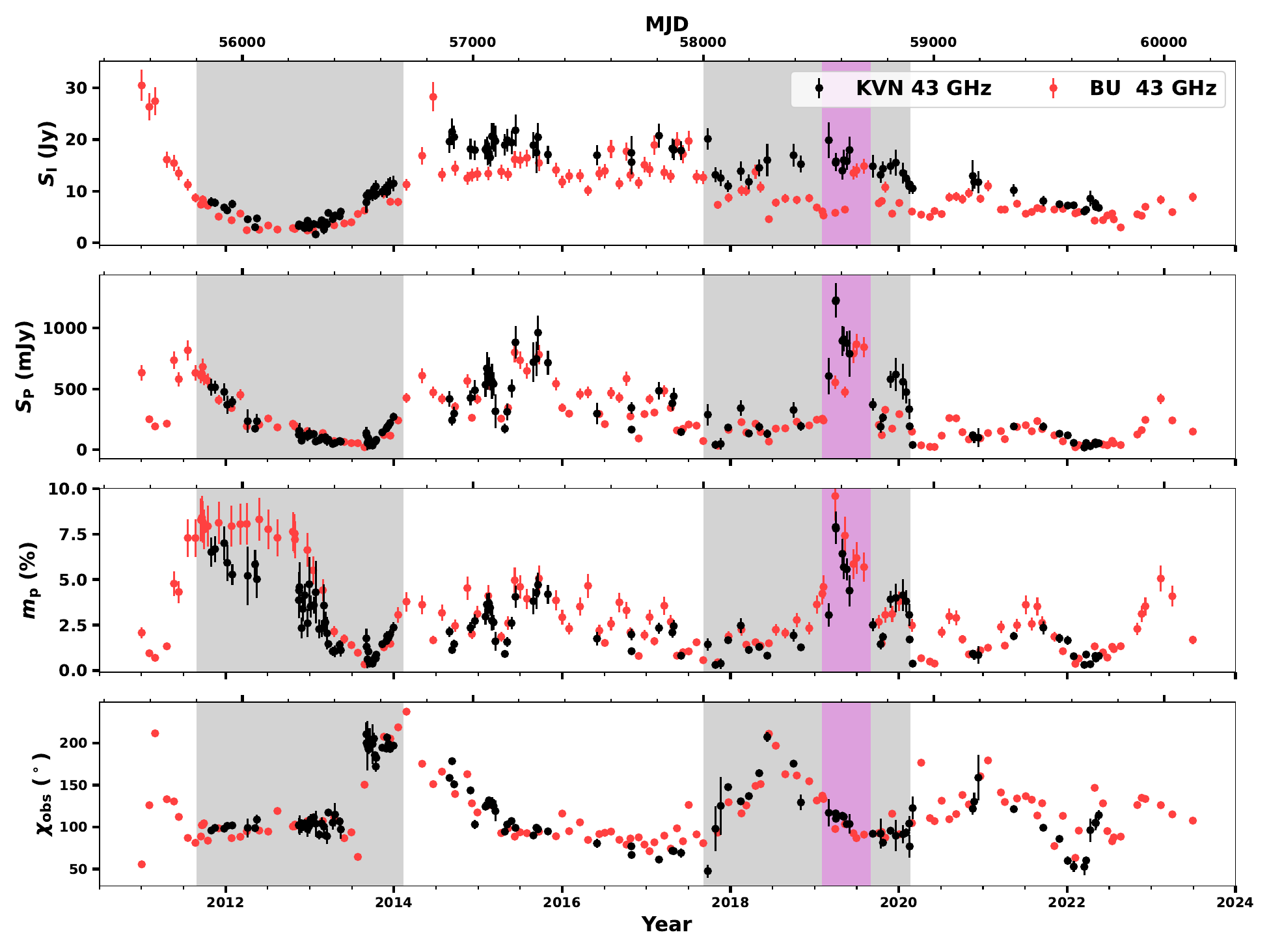}
    \caption{Comparison of the light curves at 43~GHz obtained from the KVN single-dish~(black) and the VLBA~(red). The flux density from the VLBA observations was estimated using the CLEAN models.}
    \label{fig:lc_kvn-bu}
    \end{figure}

    \FloatBarrier
    The main beam sizes of the KVN single-dish and of the VLBA are largely different. This possibly causes the additional detection of extended jet structures in the KVN observations, leading to differences in polarization characteristics. To investigate, we checked the consistency between the two data sets. To do that, we compared the light curves obtained from the KVN and the VLBA at 43~GHz~(Figure \ref{fig:lc_kvn-bu}). Each panel shows the observed values as in Figure \ref{fig:lc_kvn}. The total flux density $\Si$ of the KVN is generally higher than that of the VLBA~(by a factor of $1.49\pm0.07$). However, the polarized flux density $\Sp$ is notably comparable~(by a factor of $1.02\pm0.07$). This indicates that the extended jet beyond the VLBA main beam size has a weak or negligible impact on $\Sp$. Polarization angle depends on the ratio of the Stokes $U$ and $Q$~($\chi=0.5{\rm tan}^{-1}(U/Q)$), so $\chiobs$ may differ if there is a discrepancy between the Stokes parameters, even when the polarized flux density is comparable. From the comparison of the $\chiobs$ measurements, we found consistency between the two instruments. Therefore, we conclude that the obtained polarimetric features arise from the region within mas-scale with negligible impact from the extended downstream jet. The polarization degree $\Mp$ is slightly lower in the KVN data due to its high $\Si$.
    
    \FloatBarrier

\section{Polarization angle and rotation measure of multi-wavelength-available epochs}
\label{sec:B:appendix_rm}
    We note that all Figures in Appendix B are available on \href{10.5281/zenodo.14967589}{10.5281/zenodo.14967589}. Figure~B.1 presents the results of only 4-frequency-available epochs. In the first~(flux density, Jy) and second~(spectral index, $S_{\nu} \propto \nu^{\alpha}$) columns, the values of total and polarized emission densities are indicated by filled and open circles, respectively. The red dashed line in the second column indicates where $\alpha=0$. In the third column, the polarization degree is shown on a fixed scale~(0--12) in units of percent~($\%$). In the fourth column, the obtained polarization angle~($\chiobs$) is shown with the RM-fit~($\chi_{0}=\chi + {\rm RM}\,\lambda^{2}$) result as the red solid line in the squared wavelength~($\lambda^{2}$) domain. The calculated RM value is noted in units of $10^{3}~{\rm rad\,m^{-2}}$ in each panel. Although there are several epochs in which the linear fit well describes $\chiobs$~(i.e., constant RM), for some epochs, it cannot fit the observations. To investigate a possible frequency dependency of $\absrm$, we calculated RM at each adjacent frequency pair, as shown in the last column. The frequency in that column is the mean frequency of the pair~($\nu_{\rm mean}$). The red solid line indicates the fit result of $\absrm \propto \nu^{a}$, and the resultant $a$ index is shown in each panel. The grey shaded area indicates the range of fitted RM in absolute value~(i.e., $\rm |RM_{fit}|$). Not only the full-frequency-available $\absrm$ but also three-frequency-available results are presented in Figure~B.2. In many of the epochs, $\absrm$ seems to increase as the pair frequency (i.e., $\nu_{\rm mean}$) increases.
    \FloatBarrier
    
    \begin{table*}
    \centering
    \caption{Observed polarization degrees and polarization angles from the KVN single-dish shown in Figure~B.1.}
    \begin{tabular}{c c c c c c c c c}
    \hline\hline
        Date & \multicolumn{4}{c}{$\Mp~(\%)$} & \multicolumn{4}{c}{$\chiobs~(\degree)$}\\
    
             & 22~GHz & 43~GHz & 86~GHz & 129~GHz & 22~GHz & 43~GHz & 86~GHz & 129~GHz \\
    \hline
        2013-02-21 & $ 2.0\pm 0.2$ & $ 2.3\pm 0.4$ & $ 2.3\pm 0.2$ & $ 0.6\pm 0.5$ & $104.9\pm  3.0$ & $104.2\pm  4.1$ & $135.8\pm  4.3$ & $107.7\pm 11.7$ \\
    2013-03-03 & $ 1.8\pm 0.2$ & $ 3.6\pm 1.2$ & $ 1.5\pm 0.2$ & $ 0.3\pm 0.4$ & $105.1\pm  1.6$ & $100.4\pm  3.0$ & $126.4\pm  4.5$ & $129.5\pm 10.1$ \\
    2013-03-09 & $ 1.5\pm 0.3$ & $ 2.7\pm 0.5$ & $ 1.7\pm 0.4$ & $ 0.7\pm 1.5$ & $106.0\pm  3.6$ & $ 90.2\pm  4.1$ & $129.8\pm 10.1$ & $121.4\pm 20.5$ \\
    2013-03-16 & $ 1.4\pm 0.7$ & $ 2.0\pm 1.0$ & $ 0.3\pm 0.2$ & $ 1.3\pm 0.4$ & $104.1\pm  9.6$ & $ 89.0\pm  9.1$ & $ 99.0\pm 12.5$ & $117.0\pm 12.1$ \\
    2015-10-31 & $ 2.5\pm 0.2$ & $ 4.2\pm 0.2$ & $ 4.7\pm 0.1$ & $ 4.9\pm 0.5$ & $ 94.3\pm  1.5$ & $ 94.8\pm  0.5$ & $ 80.8\pm  0.4$ & $ 94.2\pm  2.4$ \\
    2016-05-31 & $ 1.7\pm 0.2$ & $ 1.8\pm 0.3$ & $ 4.2\pm 1.4$ & $ 4.2\pm 2.7$ & $ 91.5\pm  1.4$ & $ 80.5\pm  2.6$ & $ 77.7\pm 20.8$ & $ 78.4\pm 27.7$ \\
    2016-10-28 & $ 1.8\pm 0.1$ & $ 1.1\pm 0.1$ & $ 1.3\pm 0.1$ & $ 1.3\pm 0.2$ & $ 78.0\pm  0.9$ & $ 66.7\pm  0.9$ & $ 66.8\pm  2.9$ & $ 63.6\pm  5.0$ \\
    2017-02-24 & $ 1.9\pm 0.1$ & $ 2.3\pm 0.1$ & $ 1.8\pm 0.9$ & $ 1.9\pm 0.4$ & $ 67.1\pm  0.8$ & $ 61.2\pm  1.1$ & $ 31.9\pm 13.2$ & $ 50.6\pm  6.7$ \\
    2017-09-24 & $ 1.0\pm 0.4$ & $ 1.4\pm 0.3$ & $ 1.8\pm 1.2$ & $ 2.8\pm 2.4$ & $ 67.0\pm  7.7$ & $ 47.5\pm  7.0$ & $ 63.2\pm 27.6$ & $ 44.5\pm 23.9$ \\
    2017-10-27 & $ 0.5\pm 0.1$ & $ 0.3\pm 0.3$ & $ 1.1\pm 0.2$ & $ 1.4\pm 0.1$ & $ 85.4\pm  1.2$ & $ 97.9\pm 19.8$ & $154.8\pm  5.3$ & $120.7\pm  3.6$ \\
    2017-11-19 & $ 0.3\pm 0.2$ & $ 0.4\pm 0.4$ & $ 0.8\pm 0.6$ & $ 2.9\pm 0.9$ & $175.1\pm 16.4$ & $125.2\pm 25.5$ & $129.6\pm 25.3$ & $128.4\pm 12.8$ \\
    2018-11-02 & $ 1.2\pm 0.1$ & $ 1.3\pm 0.1$ & $ 2.9\pm 0.9$ & $ 2.5\pm 1.6$ & $166.5\pm  3.5$ & $129.3\pm  8.4$ & $152.1\pm 19.6$ & $118.8\pm 20.9$ \\
    2019-03-03 & $ 3.4\pm 0.4$ & $ 3.1\pm 0.5$ & $ 3.7\pm 0.8$ & $ 5.0\pm 0.7$ & $124.4\pm 21.2$ & $116.9\pm 12.9$ & $108.2\pm 12.4$ & $112.2\pm  3.9$ \\
    2019-04-02 & $ 5.7\pm 0.1$ & $ 7.9\pm 0.1$ & $ 9.1\pm 0.5$ & $10.3\pm 0.9$ & $116.0\pm  0.7$ & $116.2\pm  0.9$ & $ 92.4\pm  1.1$ & $112.6\pm  2.0$ \\
    2019-04-03 & $ 5.8\pm 0.1$ & $ 7.8\pm 0.1$ & $ 8.1\pm 0.5$ & $10.7\pm 0.8$ & $113.8\pm  1.5$ & $109.9\pm  1.9$ & $118.5\pm  1.4$ & $111.7\pm  2.1$ \\
    2019-05-01 & $ 5.3\pm 0.2$ & $ 6.4\pm 0.3$ & $ 7.8\pm 0.1$ & $ 9.7\pm 0.4$ & $117.2\pm  2.3$ & $113.4\pm  1.9$ & $110.7\pm  4.1$ & $104.7\pm  3.4$ \\
    2019-05-20 & $ 4.4\pm 0.1$ & $ 5.6\pm 0.1$ & $ 4.3\pm 0.6$ & $ 6.0\pm 0.6$ & $114.1\pm  0.4$ & $103.5\pm  0.7$ & $ 94.3\pm  3.9$ & $101.2\pm  3.0$ \\
    2019-06-01 & $ 3.9\pm 0.3$ & $ 4.4\pm 0.6$ & $ 5.3\pm 1.2$ & $ 5.7\pm 1.4$ & $105.7\pm  4.9$ & $103.6\pm  8.7$ & $ 90.9\pm  7.0$ & $109.1\pm  4.0$ \\
    2019-10-14 & $ 1.6\pm 0.1$ & $ 1.4\pm 0.2$ & $ 1.5\pm 0.6$ & $ 2.3\pm 1.1$ & $110.8\pm 14.2$ & $ 92.1\pm 17.1$ & $ 99.4\pm 10.9$ & $ 86.0\pm 11.0$ \\
    2019-11-26 & $ 2.3\pm 0.3$ & $ 3.9\pm 0.1$ & $ 4.7\pm 0.7$ & $ 5.1\pm 1.6$ & $101.7\pm 11.2$ & $ 95.6\pm  3.0$ & $ 97.0\pm  5.2$ & $116.7\pm  8.2$ \\
    2019-12-19 & $ 2.3\pm 0.2$ & $ 4.0\pm 0.6$ & $ 3.7\pm 0.4$ & $ 4.9\pm 0.4$ & $ 97.2\pm  8.6$ & $ 89.6\pm 18.6$ & $105.6\pm  3.3$ & $112.5\pm  5.9$ \\
    2020-01-19 & $ 2.7\pm 0.6$ & $ 4.1\pm 0.7$ & $ 4.1\pm 0.5$ & $ 4.6\pm 0.5$ & $115.7\pm 16.3$ & $ 91.3\pm 10.7$ & $100.1\pm  3.4$ & $106.8\pm  3.5$ \\
    2020-02-02 & $ 2.8\pm 0.3$ & $ 3.8\pm 0.4$ & $ 4.5\pm 0.8$ & $ 5.1\pm 1.0$ & $112.6\pm  8.8$ & $ 93.3\pm  4.2$ & $ 99.8\pm  5.0$ & $105.5\pm  3.4$ \\
    2020-11-23 & $ 0.7\pm 0.1$ & $ 0.9\pm 0.3$ & $ 2.0\pm 0.4$ & $ 1.8\pm 0.7$ & $117.0\pm 21.9$ & $130.2\pm  8.5$ & $116.9\pm  3.9$ & $119.4\pm 13.4$ \\
    2020-12-11 & $ 0.3\pm 0.4$ & $ 0.9\pm 0.6$ & $ 1.8\pm 0.4$ & $ 2.6\pm 0.9$ & $148.3\pm 49.2$ & $158.8\pm 16.1$ & $114.9\pm  5.3$ & $107.1\pm  9.5$ \\
    2021-05-14 & $ 1.5\pm 0.1$ & $ 2.0\pm 0.1$ & $ 0.7\pm 0.3$ & $ 1.6\pm 0.9$ & $132.2\pm  2.6$ & $121.3\pm  0.7$ & $ 90.9\pm 13.8$ & $ 73.3\pm 11.3$ \\
    2021-09-19 & $ 4.7\pm 3.0$ & $ 2.4\pm 0.2$ & $ 1.6\pm 0.6$ & $ 3.1\pm 2.6$ & $ 86.5\pm 13.7$ & $ 99.1\pm  2.6$ & $ 88.7\pm 10.2$ & $ 97.5\pm 24.7$ \\
    2022-01-29 & $ 0.9\pm 0.1$ & $ 0.8\pm 0.1$ & $ 1.5\pm 0.5$ & $ 2.0\pm 0.9$ & $ 55.8\pm 10.5$ & $ 52.9\pm  5.0$ & $ 30.8\pm  7.0$ & $ 47.5\pm 23.5$ \\
    2022-04-11 & $ 0.8\pm 0.1$ & $ 0.4\pm 0.1$ & $ 2.2\pm 0.8$ & $ 1.6\pm 2.5$ & $ 65.9\pm 13.9$ & $ 96.2\pm 11.0$ & $168.3\pm  8.1$ & $107.8\pm 34.3$ \\
    2022-05-03 & $ 1.0\pm 0.1$ & $ 0.8\pm 0.1$ & $ 1.6\pm 0.3$ & $ 2.0\pm 1.7$ & $ 86.3\pm 12.5$ & $105.2\pm  5.5$ & $122.7\pm  1.8$ & $115.9\pm 34.9$ \\
    2022-05-18 & $ 1.0\pm 0.1$ & $ 0.8\pm 0.1$ & $ 1.9\pm 0.6$ & $ 2.1\pm 1.3$ & $ 79.2\pm 12.3$ & $114.2\pm  2.0$ & $139.1\pm  8.2$ & $140.9\pm 20.3$ \\
    \hline
    \end{tabular}
    \label{tab:kvn_pol_result}
    \end{table*}

    \begin{table*}
    \centering
    \caption{Obtained values of the Faraday rotation measure and $a$ index. The selected epochs are the same as those listed in Table~\ref{tab:kvn_pol_result}}
    \begin{tabular}{c c c c c c}
    \hline\hline
        Date & \multicolumn{4}{c}{$\rm RM~(10^{3}~rad\,m^{-2})$} & $a~(\absrm\propto\nu^{a})$ \\
             & $\rm RM_{fit}$ & 22--43~GHz & 43--86~GHz & 86--129~GHz  & \\
    \hline
    2013-02-21 & $-1.42 \pm  1.99$ & $  0.1\pm  0.6$ & $-15.1\pm  2.8$ & $ 72.7\pm 32.3$ & $3.96 \pm 0.98$ \\
    2013-03-03 & $-2.00 \pm  1.64$ & $  0.6\pm  0.4$ & $-12.4\pm  2.6$ & $ -8.0\pm 28.6$ & $3.30 \pm 1.22$ \\
    2013-03-09 & $-1.54 \pm  2.32$ & $  2.0\pm  0.7$ & $-18.9\pm  5.2$ & $ 21.7\pm 59.1$ & $3.09 \pm 0.45$ \\
    2013-03-16 & $-0.22 \pm  1.70$ & $  1.9\pm  1.7$ & $ -4.8\pm  7.4$ & $-46.6\pm 45.0$ & $2.23 \pm 1.00$ \\
    2015-10-31 & $ 0.57 \pm  0.92$ & $ -0.1\pm  0.2$ & $  6.7\pm  0.3$ & $-34.6\pm  6.2$ & $3.77 \pm 0.76$ \\
    2016-05-31 & $ 1.33 \pm  0.09$ & $  1.4\pm  0.4$ & $  1.3\pm 10.1$ & $ -1.9\pm 89.7$ & $0.76 \pm 0.63$ \\
    2016-10-28 & $ 1.29 \pm  0.19$ & $  1.4\pm  0.2$ & $ -0.0\pm  1.5$ & $  8.3\pm 15.0$ & $0.75 \pm 1.52$ \\
    2017-02-24 & $ 2.33 \pm  1.56$ & $  0.7\pm  0.2$ & $ 14.0\pm  6.4$ & $-48.2\pm 38.3$ & $3.62 \pm 0.47$ \\
    2017-09-24 & $ 1.46 \pm  1.29$ & $  2.5\pm  1.3$ & $ -7.5\pm 13.6$ & $ 48.5\pm 94.4$ & $2.08 \pm 0.51$ \\
    2017-10-27 & $-4.73 \pm  2.97$ & $ -1.6\pm  2.5$ & $-27.2\pm  9.8$ & $ 88.0\pm 16.5$ & $2.93 \pm 0.55$ \\
    2017-11-19 & $ 4.71 \pm  1.04$ & $  6.3\pm  3.9$ & $ -2.1\pm 17.2$ & $  3.2\pm 73.2$ & $1.12 \pm 1.38$ \\
    2018-11-02 & $ 3.31 \pm  2.15$ & $  4.7\pm  1.2$ & $-10.9\pm 10.2$ & $ 86.2\pm 74.0$ & $1.62 \pm 0.72$ \\
    2019-03-03 & $ 1.35 \pm  0.38$ & $  0.9\pm  3.2$ & $  4.2\pm  8.6$ & $-10.5\pm 33.7$ & $2.04 \pm 0.12$ \\
    2019-04-02 & $ 1.14 \pm  1.47$ & $ -0.0\pm  0.1$ & $ 11.4\pm  0.7$ & $-52.3\pm  5.9$ & $3.76 \pm 0.95$ \\
    2019-04-03 & $-0.06 \pm  0.55$ & $  0.5\pm  0.3$ & $ -4.1\pm  1.1$ & $ 17.7\pm  6.6$ & $2.99 \pm 0.10$ \\
    2019-05-01 & $ 0.91 \pm  0.43$ & $  0.5\pm  0.4$ & $  1.3\pm  2.2$ & $ 15.4\pm 13.8$ & $2.65 \pm 1.07$ \\
    2019-05-20 & $ 1.57 \pm  0.48$ & $  1.4\pm  0.1$ & $  4.4\pm  1.9$ & $-18.0\pm 12.9$ & $1.87 \pm 0.24$ \\
    2019-06-01 & $ 0.48 \pm  1.12$ & $  0.3\pm  1.3$ & $  6.1\pm  5.3$ & $-47.2\pm 20.8$ & $4.08 \pm 0.15$ \\
    2019-10-14 & $ 1.88 \pm  0.83$ & $  2.4\pm  2.8$ & $ -3.5\pm  9.7$ & $ 34.7\pm 40.0$ & $1.80 \pm 1.14$ \\
    2019-11-26 & $-0.44 \pm  1.39$ & $  0.8\pm  1.5$ & $ -0.6\pm  2.9$ & $-51.1\pm 25.1$ & $8.47 \pm 5.08$ \\
    2019-12-19 & $-0.99 \pm  1.28$ & $  1.0\pm  2.6$ & $ -7.7\pm  9.1$ & $-17.8\pm 17.5$ & $2.36 \pm 0.45$ \\
    2020-01-19 & $ 1.36 \pm  1.17$ & $  3.1\pm  2.5$ & $ -4.2\pm  5.4$ & $-17.5\pm 12.7$ & $1.00 \pm 0.76$ \\
    2020-02-02 & $ 1.10 \pm  0.92$ & $  2.5\pm  1.2$ & $ -3.1\pm  3.1$ & $-14.7\pm 15.5$ & $0.89 \pm 0.67$ \\
    2020-11-23 & $-0.26 \pm  0.91$ & $ -1.7\pm  3.0$ & $  6.4\pm  4.5$ & $ -6.6\pm 36.1$ & $1.72 \pm 0.44$ \\
    2020-12-11 & $ 3.23 \pm  2.90$ & $ -1.3\pm  6.6$ & $ 21.0\pm  8.1$ & $ 20.2\pm 28.1$ & $1.77 \pm 1.26$ \\
    2021-05-14 & $ 4.66 \pm  2.25$ & $  1.4\pm  0.3$ & $ 14.6\pm  6.6$ & $ 45.4\pm 46.2$ & $3.21 \pm 0.28$ \\
    2021-09-19 & $-0.76 \pm  0.75$ & $ -1.6\pm  1.8$ & $  5.0\pm  5.0$ & $-22.7\pm 69.1$ & $1.74 \pm 0.30$ \\
    2022-01-29 & $ 1.47 \pm  1.27$ & $  0.4\pm  1.5$ & $ 10.6\pm  4.1$ & $-43.1\pm 63.4$ & $3.57 \pm 0.74$ \\
    2022-04-11 & $-6.64 \pm  4.22$ & $ -3.9\pm  2.3$ & $-34.5\pm  6.5$ & $156.4\pm 91.1$ & $3.13 \pm 0.09$ \\
    2022-05-03 & $-3.16 \pm  0.69$ & $ -2.4\pm  1.7$ & $ -8.4\pm  2.8$ & $ 17.6\pm 90.4$ & $1.82 \pm 0.04$ \\
    2022-05-18 & $-5.85 \pm  0.89$ & $ -4.4\pm  1.6$ & $-11.9\pm  4.0$ & $ -4.5\pm 56.5$ & $1.37 \pm 0.26$ \\
    \hline
    \end{tabular}
    \label{tab:kvn_rm_result}
    \end{table*}
    \FloatBarrier

\clearpage
\section{VLBA 43~GHz images during polarization flare}
\label{sec:C:appendix_polimgs}

\begin{figure}[h]
\centering
\includegraphics[width=0.95\linewidth]{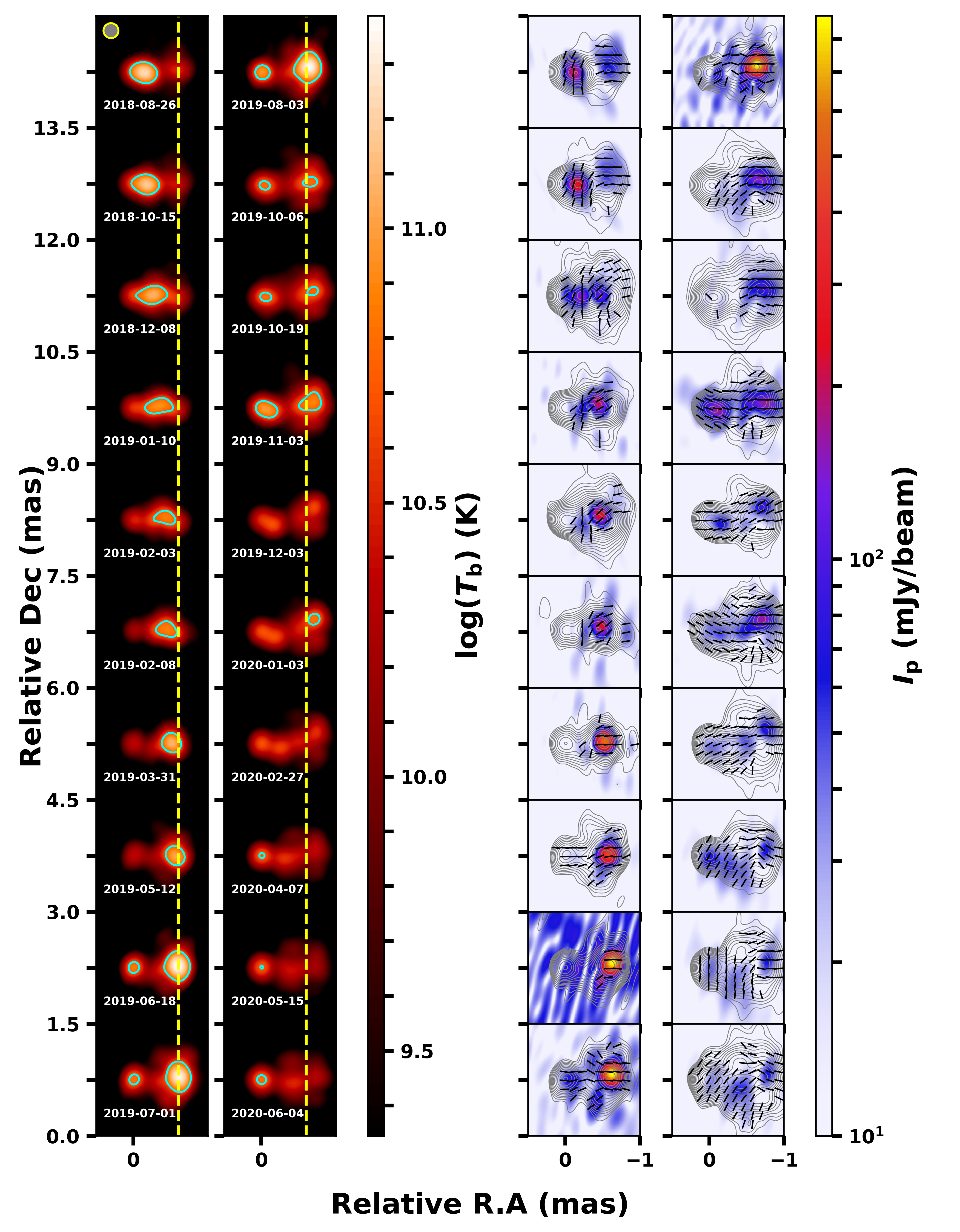}
\caption{Same as in Figure~\ref{fig:tb_2013}, but the period is around the flare in 2019.}
\label{fig:tb_2019flare}
\end{figure}

\end{appendix}

\end{document}